\documentclass[%
 reprint,
superscriptaddress,
longbibliography,
 amsmath,amssymb,
prb,
]{revtex4-1}

\bibpunct{[}{]}{;}{n}{}{}

\usepackage{dcolumn}
\usepackage{bm}

\usepackage[pdftex]{graphicx}
\usepackage{epstopdf} 
\usepackage{hyperref}
\usepackage{siunitx}
\usepackage{bm}
\usepackage{verbatim}
\usepackage[version=4]{mhchem}
\usepackage{enumitem}
\usepackage{wasysym}

\begin{document}

\preprint{APS/123-QED}

\title{Improved Single-Crystal Growth of \ce{Sr2RuO4}}

\author{J.~S.~Bobowski}
\email{jake.bobowski@ubc.ca}
\affiliation{Department of Physics, Kyoto University, Kyoto 606-8502, Japan}
\affiliation{Department of Physics, University of British Columbia, Kelowna, British Columbia, V1V 1V7, Canada}
\author{N.~Kikugawa}
\email{KIKUGAWA.Naoki@nims.go.jp}
\affiliation{National Institute for Materials Science, Tsukuba 305-0003, Japan}
\author{T.~Miyoshi}
\affiliation{Department of Physics, Kyoto University, Kyoto 606-8502, Japan}
\author{H.~Suwa}
\affiliation{Department of Physics, Kyoto University, Kyoto 606-8502, Japan}
\author{S.~Yonezawa}
\affiliation{Department of Physics, Kyoto University, Kyoto 606-8502, Japan}
\author{D.~A.~Sokolov}
\affiliation{Max Plank Institute for Chemical Physics of Solids, Dresden D-01187, Germany}
\author{A.~P.~Mackenzie}
\affiliation{Max Plank Institute for Chemical Physics of Solids, Dresden D-01187, Germany}
\author{Y.~Maeno}
\email{maeno@ss.scphys.kyoto-u.ac.jp}
\affiliation{Department of Physics, Kyoto University, Kyoto 606-8502, Japan}


\date{\today}

\begin{abstract}
High-quality single crystals are essentially needed for the investigation of the novel bulk properties of unconventional superconductors. The availability of such crystals grown by the floating-zone method has helped to unveil the unconventional superconductivity of the layered perovskite \ce{Sr2RuO4}, which is considered as a strong candidate of a topological spin-triplet superconductor. Yet, recent progress of investigations urges further efforts to obtain ultimately high-quality crystalline samples. In this paper, we focus on the method of preparation of feed rods for the floating-zone melting and report on the improvements of the crystal growth. We present details of the improved methods used to obtain crystals with superconducting transition temperatures $T_\mathrm{c}$ that are consistently as high as \SI{1.4}{\kelvin}, as well as the properties of these crystals.
\end{abstract}

\maketitle

%


\section{Introduction}
The availability of high-quality single crystals is essential for the full clarification of the novel bulk properties of quantum materials, especially of unconventional superconductors. The layered perovskite superconductor \ce{Sr2RuO4} \cite{Maeno:1994} is a typical example. This superconductor has attracted much attention over the last twenty years as a strong candidate of a spin-triplet superconductor \cite{Mackenzie:2003, Maeno:2012, Liu:2015, Kallin:2016, Mackenzie:2017}, as well as of a bulk topological superconductor \cite{Alicea:2012, Sato:2017}. Reflecting its unconventional pairing, its superconductivity is sensitively suppressed by even non-magnetic impurities \cite{Mackenzie:1998, Kikugawa:2003}. Thus, it is challenging to consistently obtain a large homogeneous crystal of \ce{Sr2RuO4} with $T_\mathrm{c}$ close to the intrinsic value of \SI{1.50}{\kelvin}, corresponding to the mean free path exceeding \SI{1}{\micro\meter}. Its superconductivity is completely suppressed when the mean-free-path becomes comparable to the superconducting coherence length ($\approx\SI{70}{\nano\meter}$), corresponding to the impurity level of c.a.\ \SI{1500}{ppm}, where the impurities act as strong scattering centers. As another characteristic of unconventional superconductivity, the quasiparticle density of states readily emerges within the superconducting gap even with a small amount of impurities and defects \cite{Dodaro:arXiv}; the resulting large residual density of states often makes the determination of the intrinsic gap anisotropy a challenging issue \cite{NishiZaki:1999, Deguchi:2004, Hassinger:2017, Kittaka:2018}. Furthermore, some unusual superconducting phenomena occur only in very high-purity crystals: the superconducting transition with a magnetic field applied parallel to the \ce{RuO2} plane becomes first order at low temperatures \cite{Yonezawa:2013, Kittaka:2014, Kikugawa:2016}. Despite the key experimental results supporting spin-triplet pairing \cite{Ishida:1998, Ishida:2000, Murakawa:2004, Duffy:2000, Ishida:2015}, such a first-order transition is difficult to explain within the context of spin-triplet superconductivity. This first-order transition becomes second order when the $T_\mathrm{c}$ is suppressed below \SI{1.45}{\kelvin}, which corresponds to an impurity level of $\sim\SI{50}{ppm}$. Thus, pristine samples with impurity levels less than \SI{10}{ppm} are required in order to deepen our knowledge of the superconducting state of \ce{Sr2RuO4}.

Recent innovations in the design of uniaxial-stress cells enables the $T_\mathrm{c}$ of \ce{Sr2RuO4} to be enhanced up to \SI{3.5}{\kelvin} when the stress is along the crystalline (100) direction \cite{Hicks:2014, Steppke:2017, Barber:2018, Luo:arXiv}. The origin of this enhancement of $T_\mathrm{c}$ is attributed to the Fermi-level crossing of the van-Hove singularity in one of the three quasi-two-dimensional Fermi surfaces. It is hoped that detailed investigations of this phenomenon will lead to the clarification of the superconducting symmetry and mechanism in \ce{Sr2RuO4}. The $T_\mathrm{c}$ enhancement has actually been known for many years in the eutectic crystals of \ce{Sr2RuO4} with micron-size metallic \ce{Ru} platelets, which introduce strong strains in \ce{Sr2RuO4} near the interfaces \cite{Maeno:1998, Ando:1999, Yaguchi:2003}. In order to investigate the strain-induced superconducting phase, high-quality \ce{Ru}-inclusion free single crystals of \ce{Sr2RuO4} with a specific in-plane crystalline direction are in demand. 

\begin{figure*}
\begin{tabular}{cccc}
(a)~\includegraphics[width=3.8 cm, keepaspectratio]{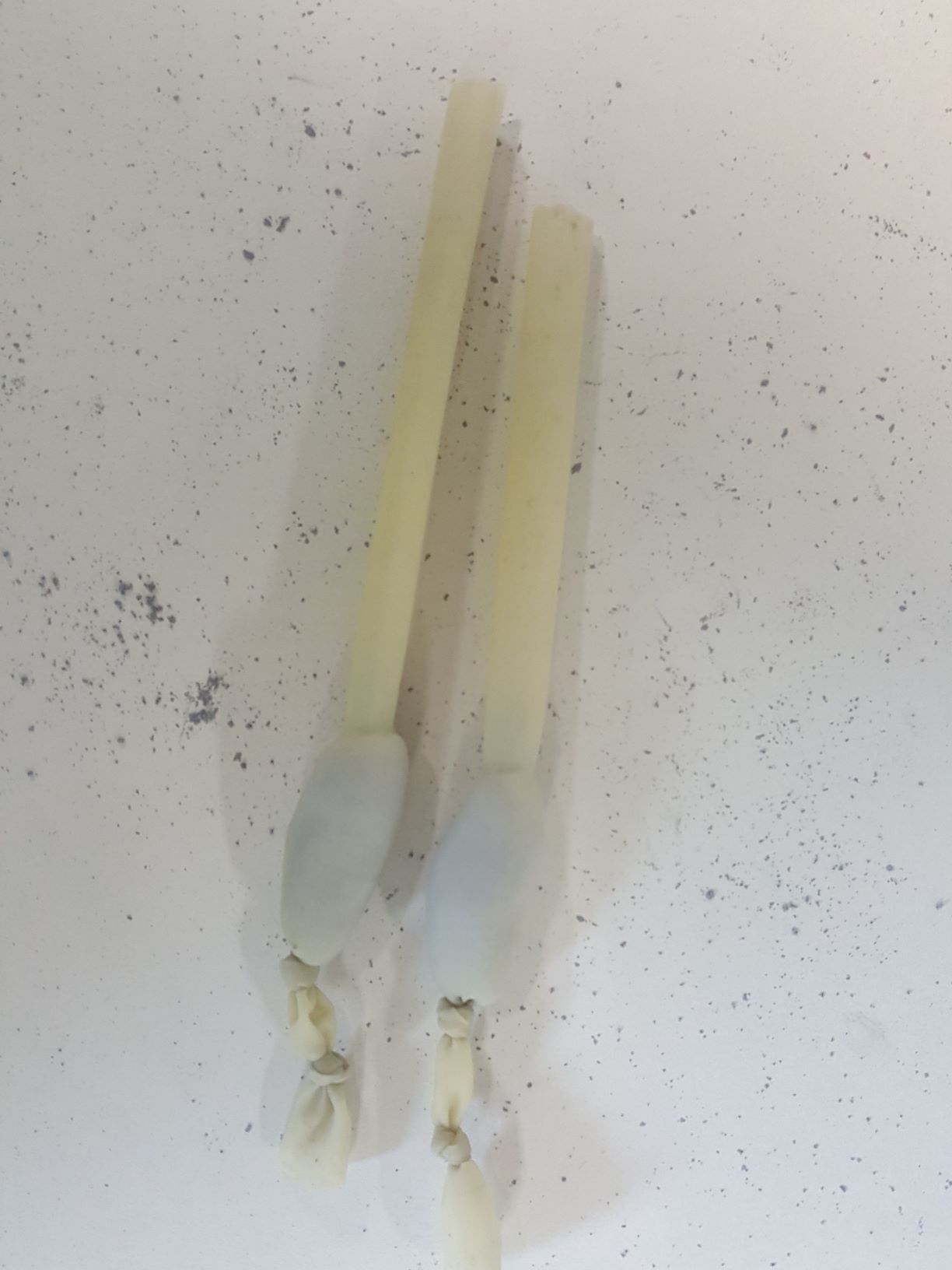} &
~(b)\includegraphics[width=3.8 cm, keepaspectratio]{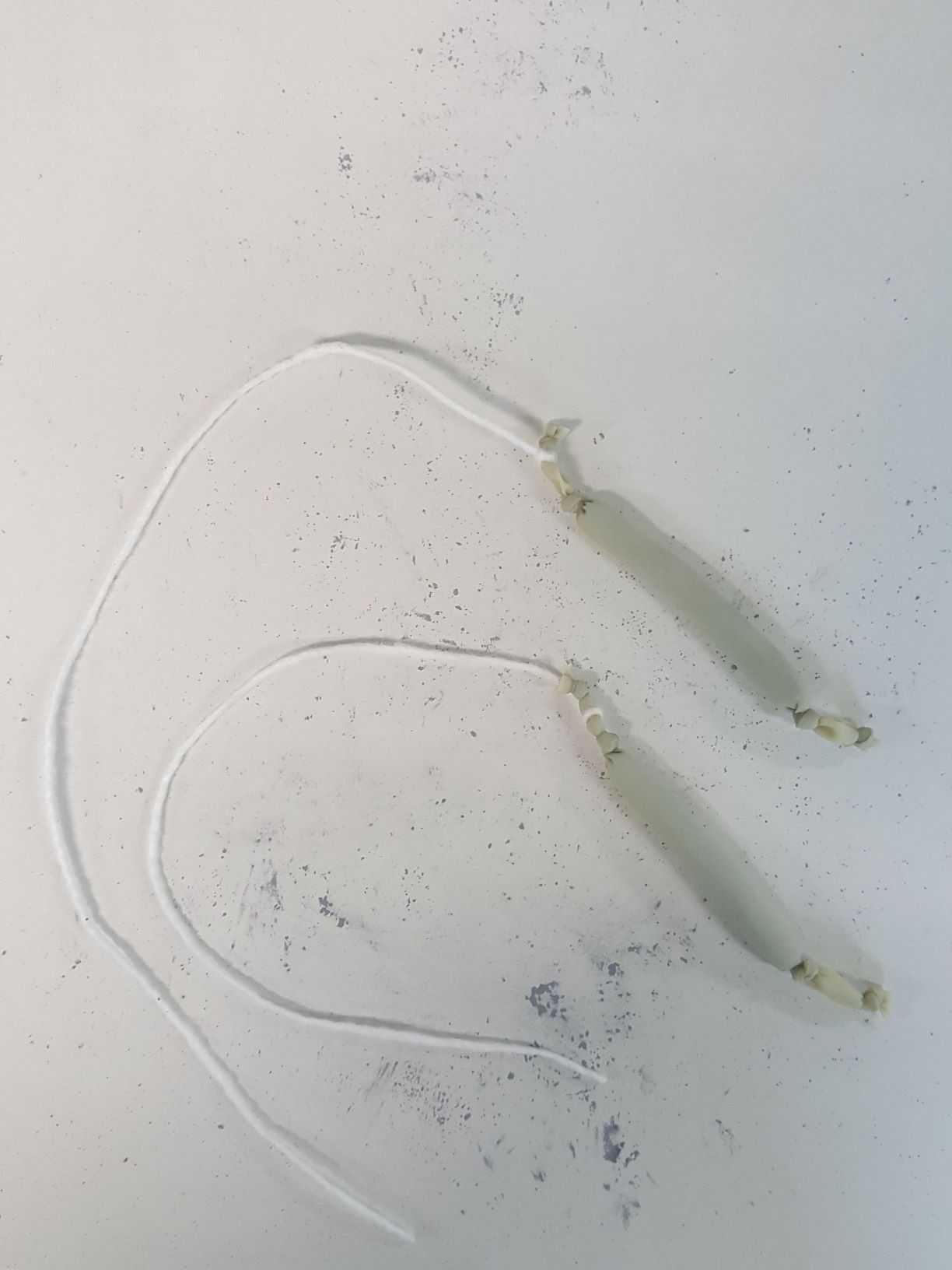} &
~(c)\includegraphics[width=3.8 cm, keepaspectratio]{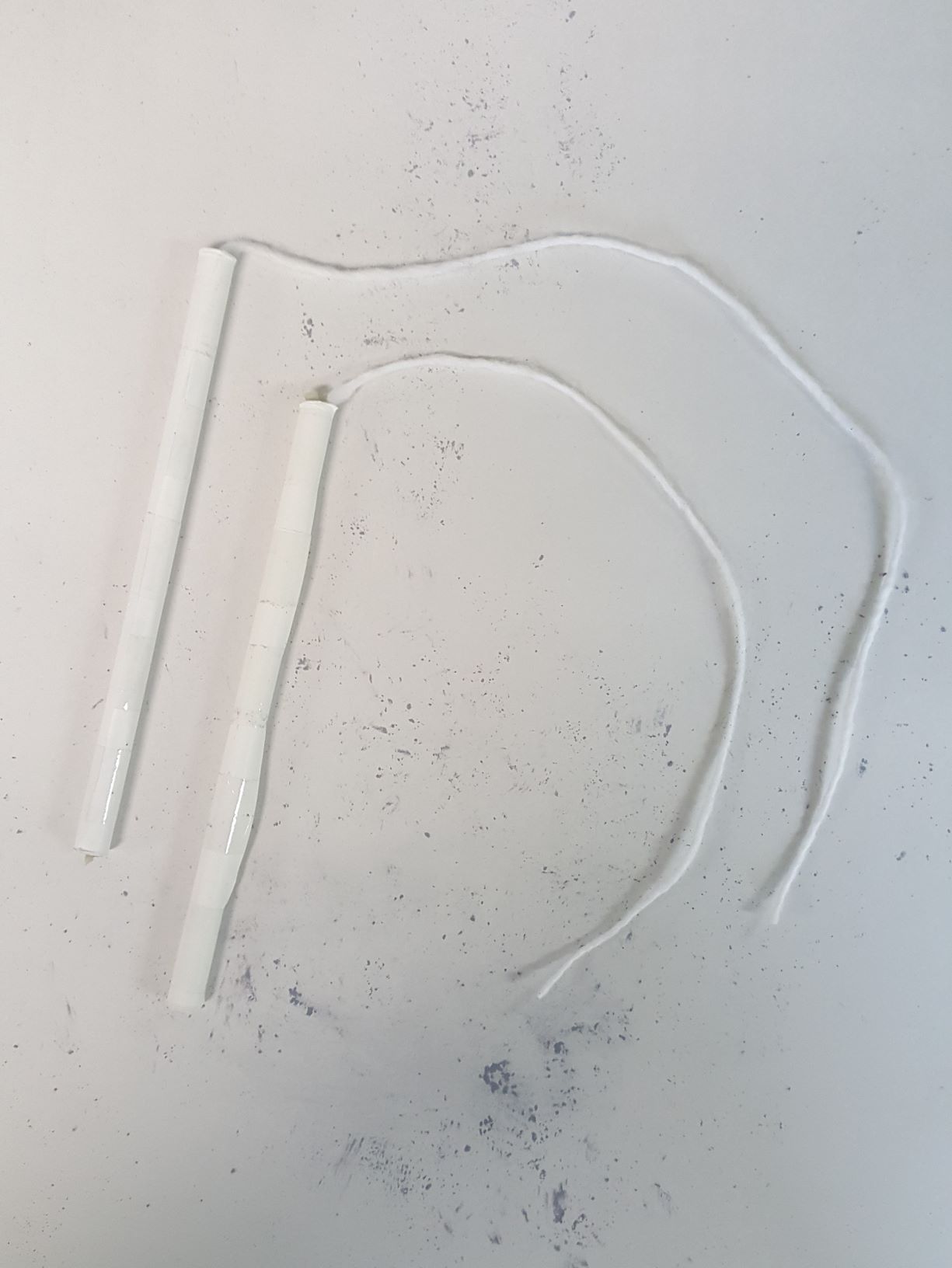} &
~(d)\includegraphics[width=3.8 cm, keepaspectratio]{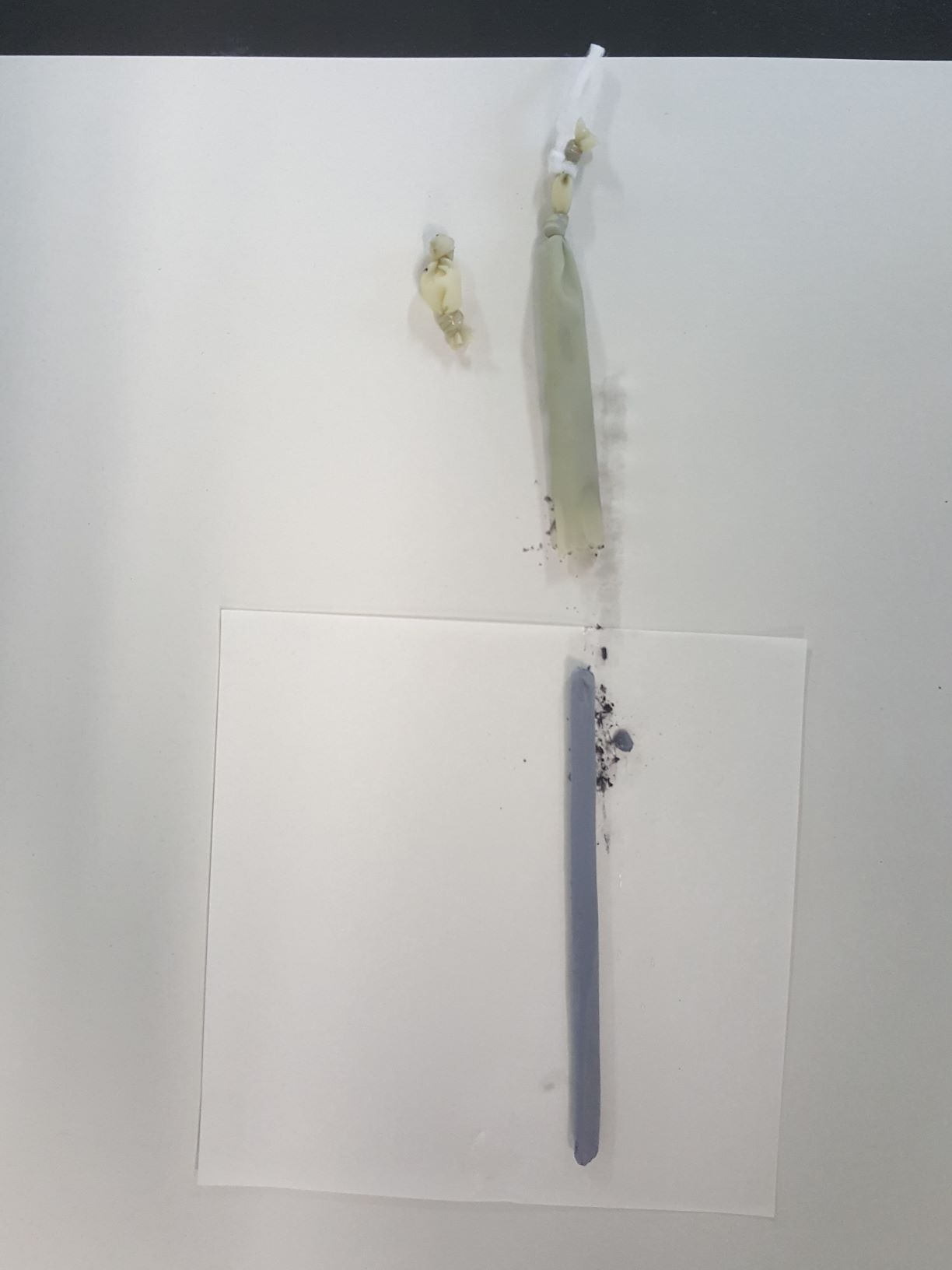}
\end{tabular}
\caption{\label{fig:01}The process of making a feed rod. (a) Balloons packed with ground \ce{SrCO3} and \ce{RuO2} powder. (b) The balloons are sealed and the powder inside is distributed as evenly as possible. A string is tied between the two knots at one end of the balloon. (c) The string is used to stretch the balloon and pull it into a paper tube with a \SI{6}{\milli\meter} inner diameter. (d) After applying hydrostatic pressure, the feed rod is removed from the balloon.}
\end{figure*}
In this study, we examine how we can further improve the quality of crystals of \ce{Sr2RuO4} grown by the floating-zone method with an infrared image furnace. The floating-zone technique has been used to produce high-quality crystals of relatively large size useful for most experimental purposes including inelastic neutron scattering. Because the technique is essentially crucible-free, it can be used to achieve the minimum possible impurity levels. The floating-zone technique has been used for the successful crystal growth of \ce{Sr2RuO4} \cite{Mao:2000}, as well as other ruthenates \cite{Perry:2004, Zhou:2005, Kikugawa:2009, Kikugawa:2015, Yoshida:2004, Kikugawa:2010}. Previous reports mainly describe optimization of the atomic compositions of the feed rod and various parameters of the final growth process. On the other hand, there are a number of processes (from powder grinding to feed-rod sintering) involved prior to the zone-melting growth process. These processes turn out to be of critical importance to minimize impurity contamination. In this study, we focus on the process of feed-rod preparation in order to further improve the quality of \ce{Sr2RuO4} crystals. Although the reexaminations of our previous processes were performed independently in Kyoto and Dresden, we achieved consistent results and reached a consistent set of conclusions. The improved feed-rod preparation processes have allowed for the efficient production of high-quality crystals which will help deepen our understanding of the superconducting state of \ce{Sr2RuO4}.

\section{Experimental}

We first describe our standard procedure of feed-rod preparation \cite{Mao:2000}, and identify possible issues. After grinding the starting materials of \ce{SrCO3} and \ce{RuO2} with the molar ratio of \ce{Sr} : \ce{Ru} = 2 : $n$ with $n = 1.15$, the powder was pressed into pellets using stainless steel molds. The pellets were typically pre-sintered at \SI{1150}{\celsius} for 24~hours. Next, the pellets were reground and the powder was packed into a balloon to shape the feed rod. For this balloon, we typically use latex tubing with a diameter of \SI{6}{\milli\meter} and a thickness of \SI{0.2}{\milli\meter}. The powder-filled balloon was then pressed under hydrostatic pressure (\SI{40}{\mega\pascal} for 5~minutes) and the resulting rod was sintered at \SI{1420}{\celsius} for 2~hours. Finally, the sintered rod was suspended in the floating-zone furnace. In this method, many steps are required before arriving at the final feed rod which allows for a number of opportunities for contamination. In particular, the surface of the pellets show a greenish color after pressing in the stainless-steel molds. This contamination is also evident from a black material on the surface of the pellets when pelletizing insulating white oxide powders instead of black \ce{RuO2}. Thus, when pellets are really needed, we cover the entire interior of the stainless-steel molds with Teflon sheets and discs. In the newly-developed processes described below, we avoid using stainless-steel molds altogether in order to establish procedures that the leave less opportunity for contamination of the feed rod. 

We have examined the following Methods (A) -- (D), in all of which we avoid using stainless-steel molds. Procedure (A) is the simplest and perhaps the least prone to contamination. Methods (B) through (D) were developed to increase the hardness of the sintered feed rod and at the same time to reduce evaporation during the melt growth.

\begin{enumerate}[label=(\Alph*)]
\item{The ground powder is packed into a balloon, then pressed, and suspended in the floating-zone furnace. Here there is no sintering process and the feed rod contains carbon from pre-reacted \ce{SrCO3}.} 
\item{The ground powder is packed into a balloon, then pressed, and sintered (\SI{1150}{\celsius} for 24~hours in Kyoto, \SI{1000}{\celsius} for 2~hours in Dresden). The sintered rod is then suspended in the floating-zone furnace.}
\item{The ground powder is packed into a balloon, then pressed, and sintered at \SI{1150}{\celsius} for 24~hours and then subsequently at \SI{1420}{\celsius} for 2~hours. The sintered rod is then suspended in the floating-zone furnace.}
\item{The ground powder is packed into a balloon, then pressed, and pre-sintered at \SI{1150}{\celsius} for 24~hours. The pre-sintered rod is reground and the powder is again packed into another balloon, then pressed, and sintered at \SI{1420}{\celsius} for 2~hours. The sintered rod is then suspended in the floating-zone furnace.}
\end{enumerate}
\begin{figure}[t]
\includegraphics[height=8.5cm, keepaspectratio]{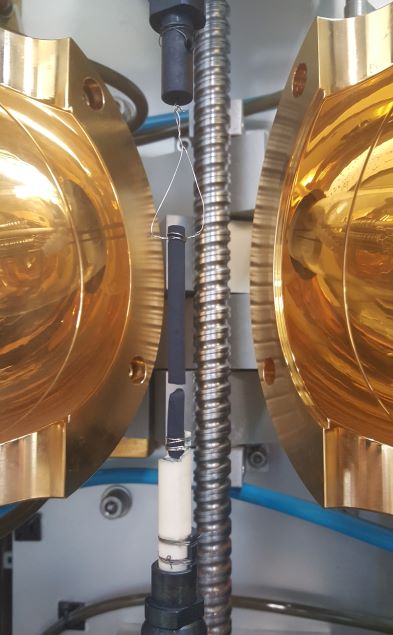}
\caption{\label{fig:02}Photo of a polycrystalline feed rod and seed ready for the floating-zone growth. They are placed at the center of the two elliptical mirrors of the infrared mirror furnace at Kyoto. In this example, a polycrystalline seed is used, rather than a single crystal.}
\end{figure}

All four procedures were examined in Kyoto, with emphasis on (C); Method (B) was adopted in Dresden. From these feed rods, \ce{Sr2RuO4} crystals were gown by the floating-zone method as described by Mao \cite{Mao:2000}. Infrared image furnaces with double-elliptical mirrors (Canon Machinery, model SC-E15HD in Kyoto, and model SC-MDH in Dresden) were used. We note that in Dresden a part of the feed rod was used as a seed during the floating-zone growth, whereas oriented single crystals or a part of the feed rod were used as seeds in Kyoto.  

We now describe the procedure used to prepare the feed rods in detail, since this is the part we particularly focus on in this study. First, two parts \ce{SrCO3} are ground with $n$ parts \ce{RuO2} in a dry nitrogen atmosphere using a mortar and pestle for at least one hour. We used 4N+ \ce{SrCO3} containing only \SI{5}{ppm} \ce{Ba}. Since \ce{SrCO3} readily absorbs moisture from the air, it is first heated to \SI{500}{\celsius} for one hour to remove adsorbed water and then weighed before being allowed to cool back to room temperature. We used 3N \ce{RuO2} typically containing a few hundred ppm \ce{Cl} and $\sim\SI{50}{ppm}$ \ce{Si} as main impurities according to the Glow Discharge Mass Spectroscopy (GDMS) analysis. The metal impurities, such as \ce{Fe}, \ce{Cu}, and \ce{Zn}, in the \ce{RuO2} are less than \SI{20}{ppm} each. Even though the analysis shows less batch dependence, we experience some batch dependence in sintering the rods; consequently, adjustment of the feed speed during the floating-zone growth is necessary. Before packing the ground powder into a latex tubing balloon, both the inner and outer surfaces of the balloon were thoroughly cleaned. The surfaces of the as-purchased latex tubing are coated with a fine \ce{TiO2} powder which prevents the balloon surfaces from sticking to itself. This powder coating must be completely removed in order to avoid unnecessary contamination of the feed rods.

Before the powder is introduced inside, a length of balloon, with both ends open, is slid over a glass rod and its outside surface is cleaned using isopropanol or ethanol and a lint-free wipe. The clean surface of the balloon is then coated with some of the ground $\ce{2SrCO3}+n\ce{RuO2}$ powder. The balloon is then inverted on the glass rod such that the opposite side can be cleaned and coated. This entire process is repeated for a second time such that the inner and outer surfaces of the balloon are each cleaned and coated twice.  

Next, one end of the balloon is tied shut with two knots separated by a few millimeters and then loaded with typically \SI{5}{\gram} of $\ce{2SrCO3}+n\ce{RuO2}$ powder (Fig.~\ref{fig:01}(a)). The powder is packed into the bottom of the balloon so as to remove as much air as possible before sealing the opposite end of the balloon with a pair of knots. Once sealed, the powder is distributed evenly throughout the length of the balloon. To form a long narrow rod, the filled balloon is inserted into a tightly-rolled tube of paper. Thick paper is rolled onto a metal rod and then secured with cellophane tape before extracting the rod. The resulting tube of paper is several layers thick and has an inner diameter of \SI{6}{\milli\meter}. A length of string is then tied between the pair of knots at one end of the balloon and used to pull the powder-filled balloon into the paper tube (Figs.~\ref{fig:01}(b) and (c)).
\begin{figure}[t]
\begin{tabular}{cc}
(a)~\includegraphics[height=7 cm, keepaspectratio]{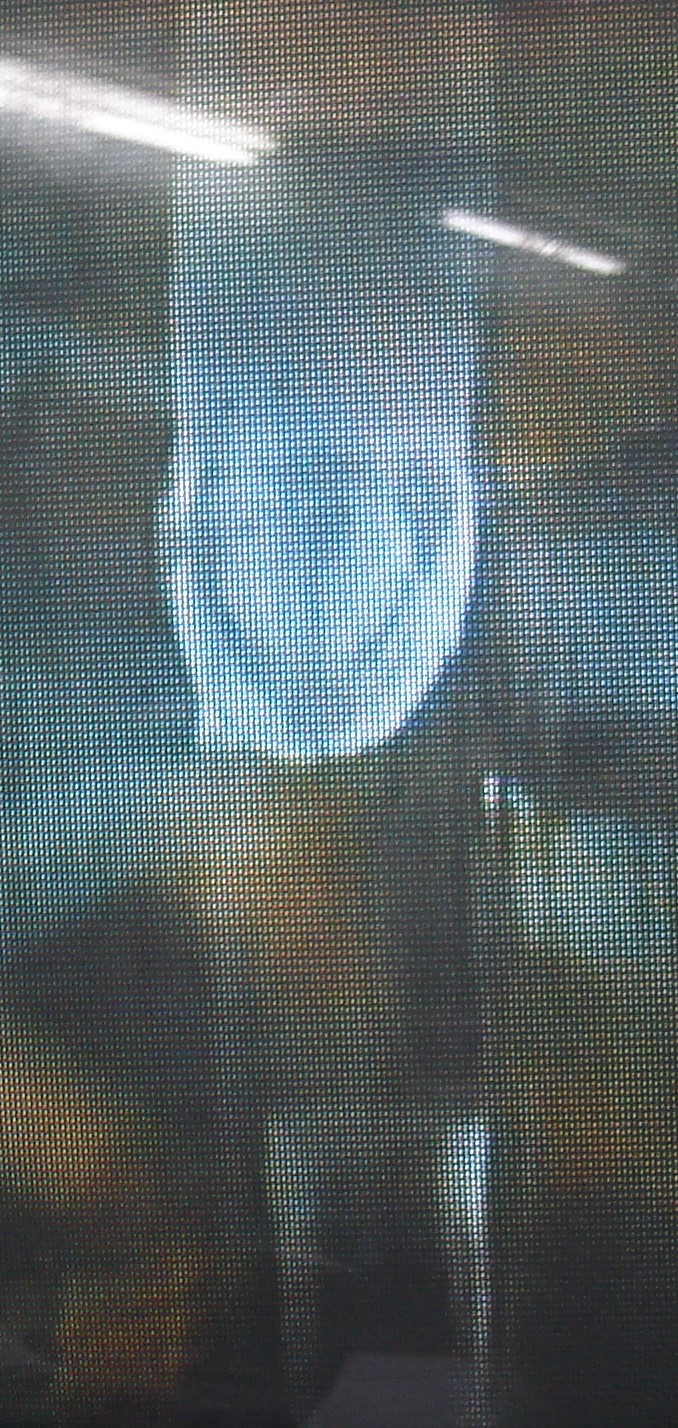} &
~(b)\includegraphics[height=7 cm, keepaspectratio]{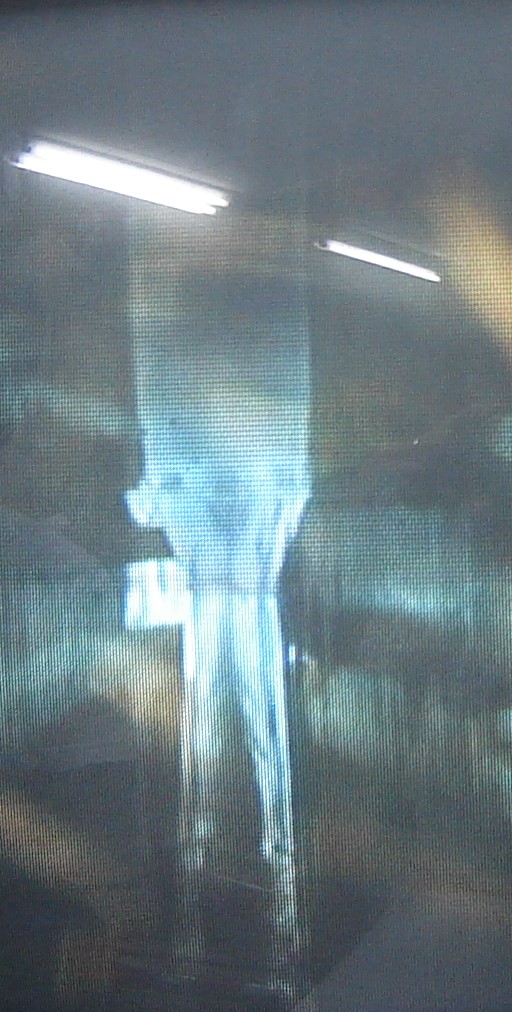}
\end{tabular}
\caption{\label{fig:03}(a) Melted tip of the feed rod just before connecting it to the seed crystal below. The diameter of the feed rod is \SI{6}{\milli\meter}. (b) The melted region between the feed rod and single crystal during stable floating-zone growth. }
\end{figure}   
\begin{figure*}
\includegraphics[width=14 cm, keepaspectratio]{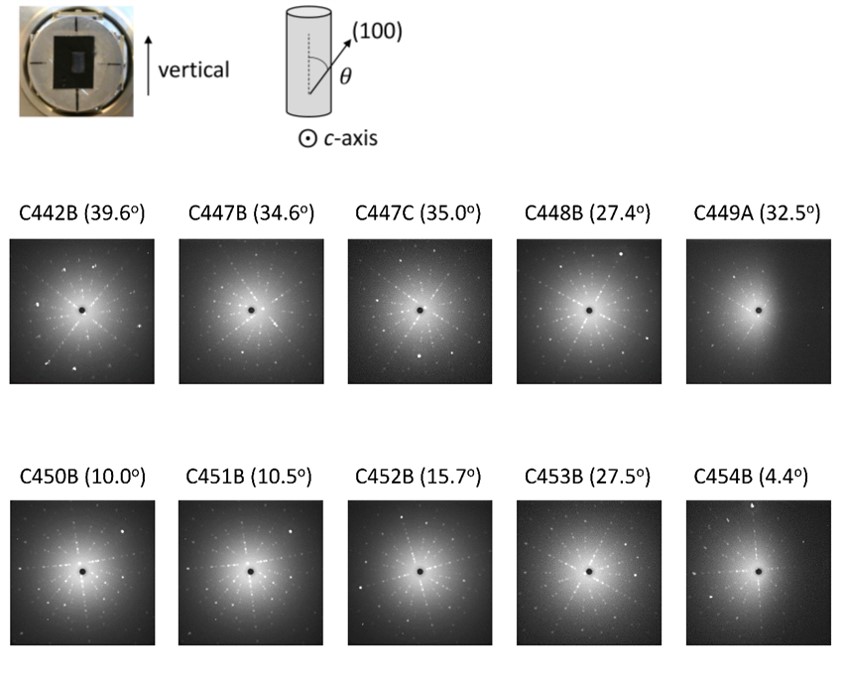}
\caption{\label{fig:04}Laue photos of the \ce{Sr2RuO4} crystals grown at Kyoto using Method (C) for the feed-rod preparation. The values in the parentheses indicate the angle of the [100] direction with respect to the crystalline-rod axis. Polycrystalline seed was used for C442, resulting in the [110] direction within 6 degrees from the crystalline-rod axis. Single-crystalline seeds were used in all other cases except C454. The same single crystal (C339) was used as seeds for C447 and C448; C441 was used as seeds for both C450 and C451, resulting in the [100] axis close to the crystalline-rod axis. For C448, crystal was disconnected during the growth and the crystal does not preserve the orientation of the seed crystal. Otherwise, the seed crystal orientations are well preserved in the crystals grown.}
\end{figure*}

The paper tube containing the powder-filled balloon is then loaded into a stainless-steel rod which has been partially bored out from one end. The bore is next filled with water and a tight-fitting piston and hydraulic press are used to apply \SI{40}{\mega\pascal} of hydrostatic pressure for five minutes. After removing and drying the paper tube, the cellophane tape is cut and the paper tube carefully uncoiled to expose the balloon. Fine scissors are used to cut one end of the balloon while one's remaining hand lightly pinches the balloon against the pressed-powder rod near the cut. After the cut is made, the pressure applied with the fingers can be slowly released in order to allow the rod to gently slide out of the balloon (Fig.~\ref{fig:01}(d)). The resulting rod, if in one piece, will be \SI{6}{\milli\meter} in diameter and up to \SI{10}{\centi\meter} long. Typically, only feed rods that are at least \SI{5}{\centi\meter} in length are suitable for floating-zone growth. The rod is very delicate and should be handled with great care.
\begin{figure*}
\begin{tabular}{lr}
(a)\includegraphics[height = 7 cm, keepaspectratio]{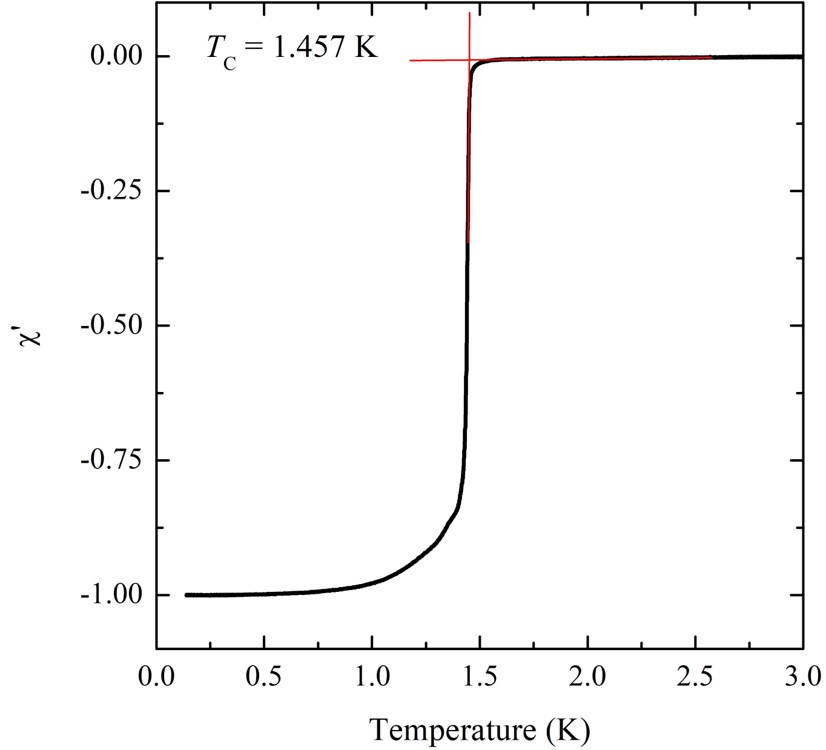}\quad~ & ~\quad(b)\includegraphics[height = 7 cm, keepaspectratio]{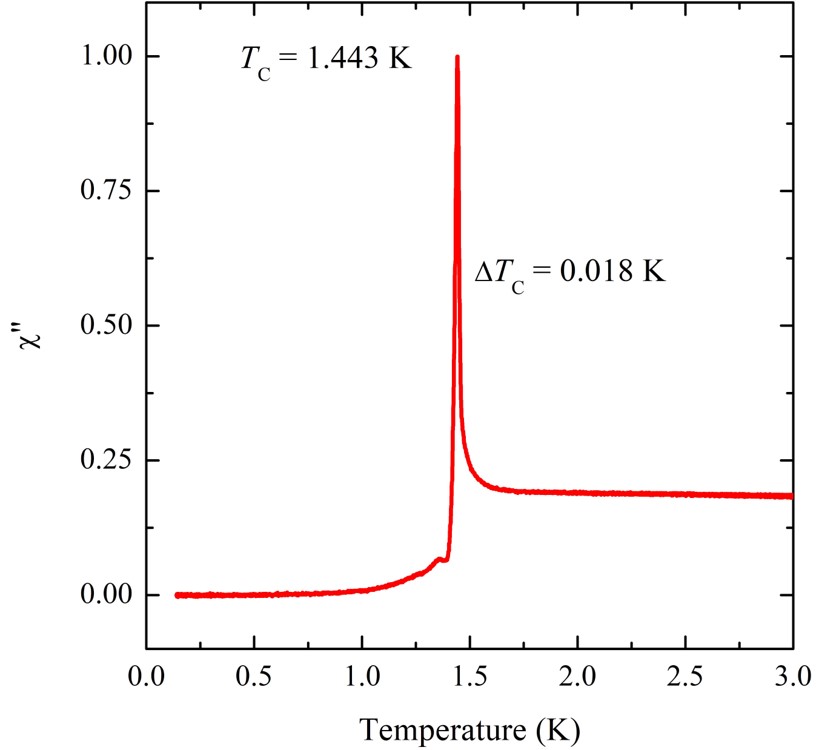}
\end{tabular}\\
\vspace{20pt}
(c)~\includegraphics[width=11 cm, keepaspectratio]{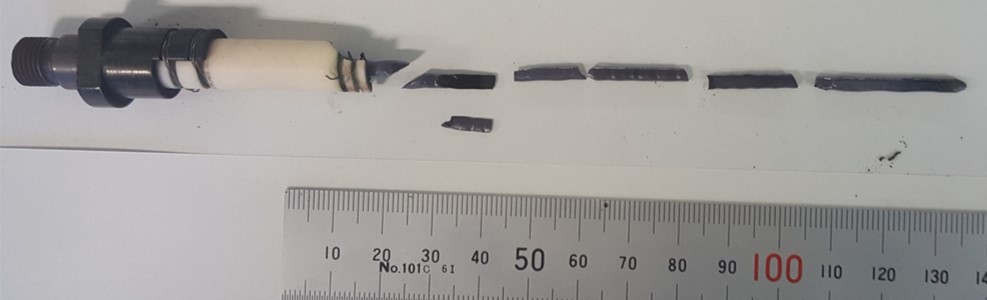}
\caption{\label{fig:05}Chi-AC and the corresponding photo of the Kyoto crystal C447B (Method (C)). (a) Real and (b) imaginary parts of the AC susceptibility of a \ce{Sr2RuO4} crystal that was prepared following Method (C). The sample that was measured was cut from the crystal shown in (c).}
\end{figure*}

In Method (A), a wire is attached to this feed rod and one then proceeds with the floating-zone growth. A utility knife can be used to carve a narrow groove near the end of the feed rod. A length of fine chromel wire (90 percent nickel and 10 percent chromium, $\diameter\,\SI{0.2}{\milli\meter}$) is wrapped around the groove in order to form a loop that can be used to suspend the feed rod in the floating-zone furnace as shown in Fig.~\ref{fig:02}. In Methods (B) -- (D), the feed rod is first sintered before the chromel wire is attached. For sintering, the feed rods are first transferred onto an alumina boat that has a bedding of pre-sintered \ce{Sr2Ru$_n$O$_{2n+2}$} powder. This bedding is used to prevent direct contact between the feed rod and alumina boat, which can be a source of unwanted contamination. In Method (B) the rods are sintered for 24~hours at \SI{1150}{\celsius} (or 2~hours at \SI{1000}{\celsius}) before the chromel wire is added. In (C) the rods are sintered at \SI{1150}{\celsius} for 24~hours followed by \SI{1420}{\celsius} for an additional 2~hours, after which the chromel wire is attached to the feed rod. Finally, in Method (D) the feed rods are first pre-sintered at \SI{1150}{\celsius} for 24~hours. These rods were then reground into powder and new feed rods were formed using the procedures described above. This second set of feed rods were then sintered at \SI{1420}{\celsius} for 2~hours before adding the chromel wire and then suspended in the floating-zone furnace. We note that the high-temperature sinter has the advantage that the resulting feed rods are less fragile and, therefore, easier to handle and manipulate. As a seed material fixed in a seed holder at the bottom (Fig.~\ref{fig:02}), we use either a part of the polycrystalline feed rod or a crystal of \ce{Sr2RuO4}.

During the floating-zone growth, some \ce{RuO2} is lost to evaporation via the reaction: 
\begin{equation}
\ce{Sr2Ru_{n}O_{$2n+2$} -> Sr2Ru_{$n^\prime$}O_{$2n^\prime+2$} + $\left(n-n^\prime\right)$RuO2}.
\label{eq:reaction1}
\end{equation}
Therefore, it is necessary to start with $n>1$ to produce single crystals with $n^\prime\cong 1$.  In Appendix~\ref{appendix} we describe how the value of $n^\prime$ can be estimated from the mass of the final single crystal after the floating-zone growth and the change in mass of the feed rod. The value of $n^\prime$ is a good indicator to monitor proper growth, and for $n = 1.15$ we obtain $n^\prime$ that are slightly less than one. So far, we have not obtained any systematic relation between $n^\prime$ and $T_\mathrm{c}$ as long as $n = 1.15$, perhaps because the precision in determining $n^\prime$ is not sufficiently high. Figure~\ref{fig:03}(a) shows photographs of a polycrystalline feed rod prepared following procedure (C) after its tip has been melted in the image furnace. The melted tip of the feed rod is suspended above a single-crystal seed.  Figure~\ref{fig:03}(b) is an image taken during stable growth of single-crystal \ce{Sr2RuO4} by the floating-zone technique. We use essentially the same growth parameters as reported previously \cite{Mao:2000}. At Kyoto, the feed speed is typically \SI{25}{\milli\meter/\hour} and crystal-growth speed is \SI{43}{\milli\meter/\hour} in a gas mixture of \ce{O2} (15\%) and \ce{Ar} (85\%) at \SI{0.25}{\mega\pascal}. The feed and seed are rotated in opposite directions, each at \SI{30}{rpm}. In Dresden, the feed speed is typically \SIrange[range-phrase=--, range-units=single]{28}{30}{\milli\meter/\hour}, and the crystal-growth speed is \SI{45}{\milli\meter/\hour} in a gas mixture of \ce{O2} (15\% or 10\%) and \ce{Ar} (85\% or 90\%) at \SI{0.35}{\mega\pascal}. The feed and seed are rotated in opposite directions typically at \SI{30}{rpm}.  

\section{Results}

Figure~\ref{fig:04} shows a series of back-scattered Laue images from \ce{Sr2RuO4} samples along the [001] direction.  All of these samples were grown using the floating-zone technique starting from feed rods prepared from Method (C). The Laue results confirm the good crystallinity of the samples. We note that high-quality samples from Method (C) have been obtained using both single crystal (C447 -- C453) and polycrystalline (C442 and C454) seeds. At Kyoto, low-temperature AC susceptibility measurements are used to assess the quality of the as-grown \ce{Sr2RuO4} crystals. High-quality samples exhibit a sharp superconducting transition with $T_\mathrm{c}$ near \SI{1.50}{\kelvin}. The AC susceptibility measurements are done using a commercial system (Quantum Design, Model PPMS) with an Adiabatic Demagnetization Refrigerator (ADR) option. The coil set and data acquisition system were designed and developed in Kyoto \cite{Yonezawa:2015}. The dimensions of the samples measured are typically $\SI{4}{\milli\meter}\times \SI{3}{\milli\meter}\times\SI{2}{\milli\meter}$ where the $c$-axis direction is along the shortest dimension and the AC magnetic field from the primary coil is applied parallel to the $ab$-plane.
\begin{figure*}
\begin{tabular}{m{7cm}m{10cm}}
	\begin{tabular}{c}
	(a)~\includegraphics[width = 6 cm, keepaspectratio]{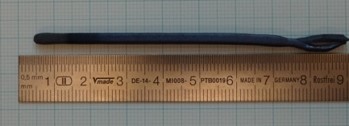}\newline\vspace{10pt}\\
	(b)~\includegraphics[width = 6 cm, keepaspectratio]{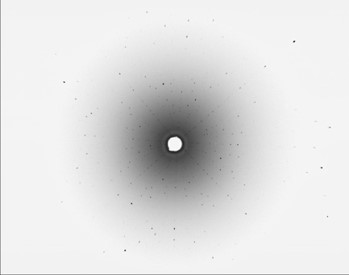}
	\end{tabular}
	& (c)\includegraphics[width = 8.5 cm, keepaspectratio]{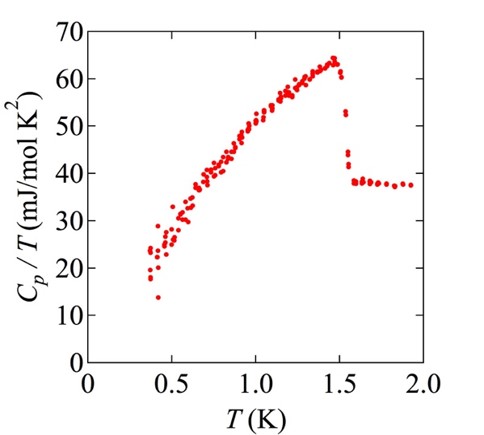}
\end{tabular}
\caption{\label{fig:06}(a) A photograph of a crystal grown by Method (B) in Dresden. (b) A back-scattered Laue-photograph along the [001] direction. (c) Specific heat ($C_p$) as a function of temperature ($T$) below \SI{2}{\kelvin}.}
\end{figure*}

Figure~\ref{fig:05} shows the real and imaginary parts of the AC susceptibility of the sample (batch-sample number C447B) whose feed rod was prepared following Method (C). For this sample, a single-crystal seed was used during the floating-zone growth. The $\chi^\prime$ data show an onset $T_\mathrm{c}$ that is greater than \SI{1.45}{\kelvin} and the width of the superconducting transition, as measured from the FWHM of $\chi^{\prime\prime}$, is less than \SI{20}{\milli\kelvin}. As described earlier, we used as a seed either a crystal of \ce{Sr2RuO4} or a part of the polycrystalline feed rod. High-quality crystals are produced in both cases. However, the samples showing sharpest superconducting transitions with transition widths less than \SI{30}{\milli\kelvin} were obtained when using single-crystal seeds. Although most of the experimental work at Kyoto has focused on Method (C), we have also successfully grown high-quality \ce{Sr2RuO4} crystals using Methods (A), (B), and (D). Method (A) has the advantage that it involves the least handling of the powders prior to the floating-zone growth and therefore is the least susceptible to contamination. However, since there is no pre-sintering the feed rods are very fragile and can be easily broken when adding the wire required to suspend the rod in the image furnace. Furthermore, these feed rods still contain carbon at the start of the floating-zone growth. Method (D) has the disadvantage that it requires the most handling of the powder prior to the floating-zone growth and these feed rods take the most time to prepare. 

A photograph of a crystal grown from a feed rod prepared by Method (B) at Dresden is shown in Fig.~\ref{fig:06}(a). Because we used a polycrystalline seed, rather than a single-crystalline seed, for this crystal, the growth at the first stage was necked. A back-scattered Laue-image from another batch of sample along the [001] direction is shown in Fig.~\ref{fig:06}(b), after cleaving the as-grown pieces along the basal plane. It is worth mentioning that crystals grown using polycrystalline seeds have a tendency of their [110] axis along the crystalline rod within \SI{10}{degrees}. This tendency has also been observed at Kyoto. We note, however, that despite this tendency there are exceptions. Sample C454 was grown at Kyoto using a polycrystalline seed and its [100] axis was found to be misaligned with the crystal-growth axis by only \SI{4.4}{\degree}. The bottom-right image in Fig.~\ref{fig:04} shows the back-scattered Laue image obtained from this sample.  In order to characterize the bulk quality of the crystal prepared following Method (B) in Dresden, we measured the specific heat ($C_p$) as a function of temperature ($T$) below \SI{2}{\kelvin} using a PPMS equipped with a $^3$He cooling system. Figure~\ref{fig:06}(c) shows the measured specific heat divided by temperature ($C_p/T$) as a function of temperature. We see a clear and sharp superconducting transition at \SI{1.5}{\kelvin}, along with a relatively small residual density of states, deduced by extrapolating $C_p/T$ to $T = \SI{0}{\kelvin}$. These observations guarantee that the grown crystal is of high quality. We note that the typical sample size used for the $C_p$ measurement is $\SI{3}{\milli\meter}\times\SI{3}{\milli\meter}\times\SI{1}{\milli\meter}$. This size is sufficient for a wide variety of experiments. 

\section{Conclusion}

We examined several feed-rod preparation procedures with the aim of avoiding impurity contamination to improve our conventional method. In particular, in all of the new procedures the feed-rod powder never comes into direct contact with any metallic surfaces. We demonstrated that the process of making pellets is unnecessary, and the crystals obtained from the feed rods prepared by the new procedures were of high quality. At Dresden, the success rate for obtaining crystals with $T_\mathrm{c} > \SI{1.45}{\kelvin}$ was 57\% among 23 different batch crystals, and at Kyoto superconducting transition widths less than \SI{20}{\milli\kelvin} as measured by AC susceptibility were obtained. The new procedures have the additional advantage that the time required to prepare the feed rods has been substantially reduced. 

For most experiments using single crystals, it is desirable to control the direction of the crystalline axes with respect to the crystalline-rod direction. For the investigation of the uniaxial-strain effects, it is desirable to have samples with the [100] direction parallel to the long axis of the crystal. In this study, we confirmed that the seed-crystal orientation is well preserved in the crystal growth and found a tendency of [110] orientation when polycrystalline feed rods are used. The improvements and additional insights acquired in this study have helped advanced us towards our goal of obtaining the ``ultimate'' crystals of \ce{Sr2RuO4} to be used for the full clarification of its superconducting state.

\vspace{10pt}
{\bf Author Contributions:} J.S.B. and Y.M. planned the project at Kyoto, whereas N.K., D.A.S, and A.P.M planned it independently at Dresden. J.S.B, T.M, H.S, S.Y. grew the crystals and characterized them at Kyoto, and N.K. and D.A.S. at Dresden. The manuscript was written mainly by Y.M., J.S.B., and N.K. with the input from all the authors. 

\vspace{10pt}
{\bf Funding:} This work was supported by JSPS KENHI Nos. JP15H05851, JP15H05852 and JP15K21717, and by JSPS
Core-to-core program Oxide Superspin. N.K. is supported by JSPS KAKENHI No. 18K04715 

\vspace{10pt}
{\bf Acknowledgments:} N.K. acknowledges Taichi Terashima, Shinya Uji, and Masahiko Kawasaki in NIMS for their support.  

\vspace{10pt}
{\bf Conflicts of Interest:} The authors declare no conflict of interest.

\appendix
\section{Estimating $n^\prime$ in \ce{Sr2Ru_{$n^\prime$}O_{$2n^\prime+2$}}}\label{appendix}

The floating-zone growth converts a polycrystalline feed rod with the composition \ce{Sr2Ru_{n}O_{$2n+2$}} into a single crystal with the composition \ce{Sr2Ru_{$n^\prime$}O_{$2n^\prime+2$}}. This appendix describes how the value of $n^\prime$ can be estimated from set of masses that can be easily measured. The estimated value of $n^\prime$ in turn helps one to optimize the value of the starting composition $n$. Before the floating-zone growth, one can measure the combined mass of the seed holder and the seed material ($m_{\mathrm{s}1}$) and the mass of the polycrystalline feed rod plus the wire that is used to suspend it in the floating-zone furnace ($m_{\mathrm{f}1}$). After the crystal growth, the total mass of the single crystal, seed and seed holder ($m_{\mathrm{s}2}$), as well as the mass of the remaining feed rod plus suspending wire ($m_{\mathrm{f}2}$) can both be measured. The difference between the final and initial seed masses results in the mass of the crystal produced during the floating-zone growth:
\begin{equation}
\Delta m_\mathrm{s}=m_{\mathrm{s}2}-m_{\mathrm{s}1}=NM_{n^\prime},\label{eq:A1}
\end{equation}
where $M_{n^\prime}$ is the molar mass of the formula unit \ce{Sr2Ru_{$n^\prime$}O_{$2n^\prime+2$}} and $N$ is the molar number. The difference in the initial and final feed rod masses specifies the total mass of polycrystalline \ce{Sr2Ru_{$n$}O_{$2n+2$}} that was converted into single crystal \ce{Sr2Ru_{$n^\prime$}O_{$2n^\prime+2$}}:
\begin{equation}
\Delta m_\mathrm{f}=m_{\mathrm{f}2}-m_{\mathrm{f}1}=-NM_n,\label{eq:A2}       
\end{equation}
where $M_n$ is the molar mass of the formula unit \ce{Sr2Ru_{$n$}O_{$2n+2$}}.  Equation~(\ref{eq:A2}) can be used to determine $N$ since $n$ is known from the initial masses of the \ce{2SrCO3} and \ce{$n$RuO2} powders (typically, $n=1.15$).  Making use of the fact that $m_n-m_{214}=(n-1)M_{\ce{RuO2}}$ allows one to write:
\begin{equation}
N=\frac{-\Delta m_f}{(n-1) M_{\ce{RuO2}} +M_{214}},\label{eq:A3}      
\end{equation}
where $M_{214}$ and $M_{\ce{RuO2}}$ are the masses of \ce{Sr2RuO4} and \ce{RuO2}, respectively.  During the single-crystal growth, the mass of \ce{RuO2} that is lost to evaporation is given by:
\begin{equation}
-\Delta m_\mathrm{f}-\Delta m_\mathrm{s}=N(n-n^\prime) M_{\ce{RuO2}}.\label{eq:A4}  
\end{equation}
Eliminating $N$ from Eqs.~(\ref{eq:A3}) and (\ref{eq:A4}) and solving for $n^\prime$ leads to:
\begin{equation}
n^\prime=n\left\vert\frac{\Delta m_\mathrm{s}}{\Delta m_\mathrm{f}}\right\vert +\left(\frac{M_{214}}{M_{\ce{RuO2}}}-1\right)\left(\left\vert\frac{\Delta m_\mathrm{s}}{\Delta m_\mathrm{f}}\right\vert-1\right),\label{eq:A5}       
\end{equation}
where the ratio $M_{214}/M_{\ce{RuO2}}=2.5574$.

\nocite{*}
\bibliography{Sr214}

\providecommand{\noopsort}[1]{}\providecommand{\singleletter}[1]{#1}%
\begin{thebibliography}{38}%
\makeatletter
\providecommand \@ifxundefined [1]{%
 \@ifx{#1\undefined}
}%
\providecommand \@ifnum [1]{%
 \ifnum #1\expandafter \@firstoftwo
 \else \expandafter \@secondoftwo
 \fi
}%
\providecommand \@ifx [1]{%
 \ifx #1\expandafter \@firstoftwo
 \else \expandafter \@secondoftwo
 \fi
}%
\providecommand \natexlab [1]{#1}%
\providecommand \enquote  [1]{``#1''}%
\providecommand \bibnamefont  [1]{#1}%
\providecommand \bibfnamefont [1]{#1}%
\providecommand \citenamefont [1]{#1}%
\providecommand \href@noop [0]{\@secondoftwo}%
\providecommand \href [0]{\begingroup \@sanitize@url \@href}%
\providecommand \@href[1]{\@@startlink{#1}\@@href}%
\providecommand \@@href[1]{\endgroup#1\@@endlink}%
\providecommand \@sanitize@url [0]{\catcode `\\12\catcode `\$12\catcode
  `\&12\catcode `\#12\catcode `\^12\catcode `\_12\catcode `\%12\relax}%
\providecommand \@@startlink[1]{}%
\providecommand \@@endlink[0]{}%
\providecommand \url  [0]{\begingroup\@sanitize@url \@url }%
\providecommand \@url [1]{\endgroup\@href {#1}{\urlprefix }}%
\providecommand \urlprefix  [0]{URL }%
\providecommand \Eprint [0]{\href }%
\providecommand \doibase [0]{http://dx.doi.org/}%
\providecommand \selectlanguage [0]{\@gobble}%
\providecommand \bibinfo  [0]{\@secondoftwo}%
\providecommand \bibfield  [0]{\@secondoftwo}%
\providecommand \translation [1]{[#1]}%
\providecommand \BibitemOpen [0]{}%
\providecommand \bibitemStop [0]{}%
\providecommand \bibitemNoStop [0]{.\EOS\space}%
\providecommand \EOS [0]{\spacefactor3000\relax}%
\providecommand \BibitemShut  [1]{\csname bibitem#1\endcsname}%
\let\auto@bib@innerbib\@empty
\bibitem [{\citenamefont {Maeno}\ \emph {et~al.}(1994)\citenamefont {Maeno},
  \citenamefont {Hashimoto}, \citenamefont {Yoshida}, \citenamefont
  {Nishizaki}, \citenamefont {Fujita}, \citenamefont {Bednorz},\ and\
  \citenamefont {Lichtenberg}}]{Maeno:1994}%
  \BibitemOpen
  \bibfield  {author} {\bibinfo {author} {\bibfnamefont {Y.}~\bibnamefont
  {Maeno}}, \bibinfo {author} {\bibfnamefont {H.}~\bibnamefont {Hashimoto}},
  \bibinfo {author} {\bibfnamefont {K.}~\bibnamefont {Yoshida}}, \bibinfo
  {author} {\bibfnamefont {S.}~\bibnamefont {Nishizaki}}, \bibinfo {author}
  {\bibfnamefont {T.}~\bibnamefont {Fujita}}, \bibinfo {author} {\bibfnamefont
  {J.~G.}\ \bibnamefont {Bednorz}}, \ and\ \bibinfo {author} {\bibfnamefont
  {F.}~\bibnamefont {Lichtenberg}},\ }\bibfield  {title} {\enquote {\bibinfo
  {title} {Superconductivity in a layered perovskite without copper},}\ }\href
  {\doibase 10.1038/372532a0} {\bibfield  {journal} {\bibinfo  {journal}
  {Nature}\ }\textbf {\bibinfo {volume} {372}},\ \bibinfo {pages} {532--534}
  (\bibinfo {year} {1994})}\BibitemShut {NoStop}%
\bibitem [{\citenamefont {Mackenzie}\ and\ \citenamefont
  {Maeno}(2003)}]{Mackenzie:2003}%
  \BibitemOpen
  \bibfield  {author} {\bibinfo {author} {\bibfnamefont {A.~P.}\ \bibnamefont
  {Mackenzie}}\ and\ \bibinfo {author} {\bibfnamefont {Y.}~\bibnamefont
  {Maeno}},\ }\bibfield  {title} {\enquote {\bibinfo {title} {The
  superconductivity of \ce{Sr2RuO4} and the physics of spin-triplet pairing},}\
  }\href {\doibase 10.1103/RevModPhys.75.657} {\bibfield  {journal} {\bibinfo
  {journal} {Rev.\ Mod.\ Phys.}\ }\textbf {\bibinfo {volume} {75}},\ \bibinfo
  {pages} {657--712} (\bibinfo {year} {2003})}\BibitemShut {NoStop}%
\bibitem [{\citenamefont {Maeno}\ \emph {et~al.}(2012)\citenamefont {Maeno},
  \citenamefont {Kittaka}, \citenamefont {Nomura}, \citenamefont {Yonezawa},\
  and\ \citenamefont {Ishida}}]{Maeno:2012}%
  \BibitemOpen
  \bibfield  {author} {\bibinfo {author} {\bibfnamefont {Y.}~\bibnamefont
  {Maeno}}, \bibinfo {author} {\bibfnamefont {S.}~\bibnamefont {Kittaka}},
  \bibinfo {author} {\bibfnamefont {T.}~\bibnamefont {Nomura}}, \bibinfo
  {author} {\bibfnamefont {S.}~\bibnamefont {Yonezawa}}, \ and\ \bibinfo
  {author} {\bibfnamefont {K.}~\bibnamefont {Ishida}},\ }\bibfield  {title}
  {\enquote {\bibinfo {title} {Evaluation of spin-triplet superconductivity in
  \ce{Sr2RuO4}},}\ }\href {\doibase 10.1143/JPSJ.81.011009} {\bibfield
  {journal} {\bibinfo  {journal} {J.\ Phys.\ Soc.\ Jpn.}\ }\textbf {\bibinfo
  {volume} {81}},\ \bibinfo {pages} {011009} (\bibinfo {year}
  {2012})}\BibitemShut {NoStop}%
\bibitem [{\citenamefont {Liu}\ and\ \citenamefont {Mao}(2015)}]{Liu:2015}%
  \BibitemOpen
  \bibfield  {author} {\bibinfo {author} {\bibfnamefont {Y.}~\bibnamefont
  {Liu}}\ and\ \bibinfo {author} {\bibfnamefont {Z.-Q.}\ \bibnamefont {Mao}},\
  }\bibfield  {title} {\enquote {\bibinfo {title} {Unconventional
  superconductivity in \ce{Sr2RuO4}},}\ }\href {\doibase
  10.1016/j.physc.2015.02.039} {\bibfield  {journal} {\bibinfo  {journal}
  {Physica C}\ }\textbf {\bibinfo {volume} {514}},\ \bibinfo {pages} {339--353}
  (\bibinfo {year} {2015})}\BibitemShut {NoStop}%
\bibitem [{\citenamefont {Kallin}\ and\ \citenamefont
  {Berlinsky}(2016)}]{Kallin:2016}%
  \BibitemOpen
  \bibfield  {author} {\bibinfo {author} {\bibfnamefont {C.}~\bibnamefont
  {Kallin}}\ and\ \bibinfo {author} {\bibfnamefont {J.}~\bibnamefont
  {Berlinsky}},\ }\bibfield  {title} {\enquote {\bibinfo {title} {Chiral
  superconductors},}\ }\href {\doibase 10.1088/0034-4885/79/5/054502}
  {\bibfield  {journal} {\bibinfo  {journal} {Rep.\ Prog.\ Phys.}\ }\textbf
  {\bibinfo {volume} {79}},\ \bibinfo {pages} {054502} (\bibinfo {year}
  {2016})}\BibitemShut {NoStop}%
\bibitem [{\citenamefont {Mackenzie}\ \emph {et~al.}(2017)\citenamefont
  {Mackenzie}, \citenamefont {Scaffidi}, \citenamefont {Hicks},\ and\
  \citenamefont {Maeno}}]{Mackenzie:2017}%
  \BibitemOpen
  \bibfield  {author} {\bibinfo {author} {\bibfnamefont {A.~P.}\ \bibnamefont
  {Mackenzie}}, \bibinfo {author} {\bibfnamefont {T.}~\bibnamefont {Scaffidi}},
  \bibinfo {author} {\bibfnamefont {C.~W.}\ \bibnamefont {Hicks}}, \ and\
  \bibinfo {author} {\bibfnamefont {Y.}~\bibnamefont {Maeno}},\ }\bibfield
  {title} {\enquote {\bibinfo {title} {Even odder after twenty-three years:
  {T}he superconducting order parameter puzzle of \ce{Sr2RuO4}},}\ }\href
  {\doibase 10.1038/s41535-017-0045-4} {\bibfield  {journal} {\bibinfo
  {journal} {npj Quantum Mater.}\ }\textbf {\bibinfo {volume} {2}},\ \bibinfo
  {pages} {40} (\bibinfo {year} {2017})}\BibitemShut {NoStop}%
\bibitem [{\citenamefont {Alicea}(2012)}]{Alicea:2012}%
  \BibitemOpen
  \bibfield  {author} {\bibinfo {author} {\bibfnamefont {J.}~\bibnamefont
  {Alicea}},\ }\bibfield  {title} {\enquote {\bibinfo {title} {New directions
  in the pursuit of majorana fermions in solid state systems},}\ }\href
  {\doibase 10.1088/0034-4885/75/7/076501} {\bibfield  {journal} {\bibinfo
  {journal} {Rep.\ Prog.\ Phys.}\ }\textbf {\bibinfo {volume} {75}},\ \bibinfo
  {pages} {076501} (\bibinfo {year} {2012})}\BibitemShut {NoStop}%
\bibitem [{\citenamefont {Sato}\ and\ \citenamefont {Ando}(2017)}]{Sato:2017}%
  \BibitemOpen
  \bibfield  {author} {\bibinfo {author} {\bibfnamefont {M.}~\bibnamefont
  {Sato}}\ and\ \bibinfo {author} {\bibfnamefont {Y.}~\bibnamefont {Ando}},\
  }\bibfield  {title} {\enquote {\bibinfo {title} {Topological superconductors:
  {A} review},}\ }\href {\doibase 10.1088/1361-6633/aa6ac7} {\bibfield
  {journal} {\bibinfo  {journal} {Rep.\ Prog.\ Phys.}\ }\textbf {\bibinfo
  {volume} {80}},\ \bibinfo {pages} {076501} (\bibinfo {year}
  {2017})}\BibitemShut {NoStop}%
\bibitem [{\citenamefont {Mackenzie}\ \emph {et~al.}(1998)\citenamefont
  {Mackenzie}, \citenamefont {Haselwimmer}, \citenamefont {Tyler},
  \citenamefont {Lonzarich}, \citenamefont {Mori}, \citenamefont {Nishizaki},\
  and\ \citenamefont {Maeno}}]{Mackenzie:1998}%
  \BibitemOpen
  \bibfield  {author} {\bibinfo {author} {\bibfnamefont {A.~P.}\ \bibnamefont
  {Mackenzie}}, \bibinfo {author} {\bibfnamefont {R.~K.~W.}\ \bibnamefont
  {Haselwimmer}}, \bibinfo {author} {\bibfnamefont {A.~W.}\ \bibnamefont
  {Tyler}}, \bibinfo {author} {\bibfnamefont {G.~G.}\ \bibnamefont
  {Lonzarich}}, \bibinfo {author} {\bibfnamefont {Y.}~\bibnamefont {Mori}},
  \bibinfo {author} {\bibfnamefont {S.}~\bibnamefont {Nishizaki}}, \ and\
  \bibinfo {author} {\bibfnamefont {Y.}~\bibnamefont {Maeno}},\ }\bibfield
  {title} {\enquote {\bibinfo {title} {Extremely strong dependence of
  superconductivity on disorder in \ce{Sr2RuO4}},}\ }\href {\doibase
  10.1103/PhysRevLett.80.161} {\bibfield  {journal} {\bibinfo  {journal}
  {Phys.\ Rev.\ Lett.}\ }\textbf {\bibinfo {volume} {80}},\ \bibinfo {pages}
  {161--164} (\bibinfo {year} {1998})}\BibitemShut {NoStop}%
\bibitem [{\citenamefont {Kikugawa}\ \emph {et~al.}(2003)\citenamefont
  {Kikugawa}, \citenamefont {Mackenzie},\ and\ \citenamefont
  {Maeno}}]{Kikugawa:2003}%
  \BibitemOpen
  \bibfield  {author} {\bibinfo {author} {\bibfnamefont {N.}~\bibnamefont
  {Kikugawa}}, \bibinfo {author} {\bibfnamefont {A.~P.}\ \bibnamefont
  {Mackenzie}}, \ and\ \bibinfo {author} {\bibfnamefont {Y.}~\bibnamefont
  {Maeno}},\ }\bibfield  {title} {\enquote {\bibinfo {title} {Effects of
  in-plane impurity substitution in \ce{Sr2RuO4}},}\ }\href {\doibase
  10.1143/JPSJ.72.237} {\bibfield  {journal} {\bibinfo  {journal} {J.\ Phys.\
  Soc.\ Jpn.}\ }\textbf {\bibinfo {volume} {72}},\ \bibinfo {pages} {237--240}
  (\bibinfo {year} {2003})}\BibitemShut {NoStop}%
\bibitem [{\citenamefont {Dodaro}\ \emph {et~al.}()\citenamefont {Dodaro},
  \citenamefont {Wang},\ and\ \citenamefont {Kallin}}]{Dodaro:arXiv}%
  \BibitemOpen
  \bibfield  {author} {\bibinfo {author} {\bibfnamefont {J.~F.}\ \bibnamefont
  {Dodaro}}, \bibinfo {author} {\bibfnamefont {Z.}~\bibnamefont {Wang}}, \ and\
  \bibinfo {author} {\bibfnamefont {C.}~\bibnamefont {Kallin}},\ }\bibfield
  {title} {\enquote {\bibinfo {title} {Effects of deep superconducting gap
  minima and disorder on residual thermal transport in \ce{Sr2RuO4}},}\ }\href
  {https://arxiv.org/abs/1810.00932} {\bibfield  {journal} {\bibinfo  {journal}
  {{\it arXiv:1810.00932v1}}\ }}\bibinfo {note}
  {{h}ttps://arxiv.org/abs/1810.00932}\BibitemShut {NoStop}%
\bibitem [{\citenamefont {NishiZaki}\ \emph {et~al.}(1999)\citenamefont
  {NishiZaki}, \citenamefont {Maeno}, ,\ and\ \citenamefont
  {Mao}}]{NishiZaki:1999}%
  \BibitemOpen
  \bibfield  {author} {\bibinfo {author} {\bibfnamefont {S.}~\bibnamefont
  {NishiZaki}}, \bibinfo {author} {\bibfnamefont {Y.}~\bibnamefont {Maeno}}, ,
  \ and\ \bibinfo {author} {\bibfnamefont {Z.-Q.}\ \bibnamefont {Mao}},\
  }\bibfield  {title} {\enquote {\bibinfo {title} {Effect of impurities on the
  specific heat of the spin-triplet superconductor \ce{Sr2RuO4}},}\ }\href
  {\doibase 10.1023/A:1022551313401} {\bibfield  {journal} {\bibinfo  {journal}
  {J.\ Low Temp.\ Phys.}\ }\textbf {\bibinfo {volume} {117}},\ \bibinfo {pages}
  {1581--1585} (\bibinfo {year} {1999})}\BibitemShut {NoStop}%
\bibitem [{\citenamefont {Deguchi}\ \emph {et~al.}(2004)\citenamefont
  {Deguchi}, \citenamefont {Mao}, \citenamefont {Yaguchi},\ and\ \citenamefont
  {Maeno}}]{Deguchi:2004}%
  \BibitemOpen
  \bibfield  {author} {\bibinfo {author} {\bibfnamefont {K.}~\bibnamefont
  {Deguchi}}, \bibinfo {author} {\bibfnamefont {Z.-Q.}\ \bibnamefont {Mao}},
  \bibinfo {author} {\bibfnamefont {H.}~\bibnamefont {Yaguchi}}, \ and\
  \bibinfo {author} {\bibfnamefont {Y.}~\bibnamefont {Maeno}},\ }\bibfield
  {title} {\enquote {\bibinfo {title} {Gap structure of the spin-triplet
  superconductor \ce{Sr2RuO4} determined from the field-orientation dependence
  of the specific heat},}\ }\href {\doibase 10.1103/PhysRevLett.92.047002}
  {\bibfield  {journal} {\bibinfo  {journal} {Phys.\ Rev.\ Lett.}\ }\textbf
  {\bibinfo {volume} {92}},\ \bibinfo {pages} {047002} (\bibinfo {year}
  {2004})}\BibitemShut {NoStop}%
\bibitem [{\citenamefont {Hassinger}\ \emph {et~al.}(2017)\citenamefont
  {Hassinger}, \citenamefont {Bourgeois-Hope}, \citenamefont {Taniguchi},
  \citenamefont {Ren\'e~de Cotret}, \citenamefont {Grissonnanche},
  \citenamefont {Anwar}, \citenamefont {Maeno}, \citenamefont
  {Doiron-Leyraud},\ and\ \citenamefont {Taillefer}}]{Hassinger:2017}%
  \BibitemOpen
  \bibfield  {author} {\bibinfo {author} {\bibfnamefont {E.}~\bibnamefont
  {Hassinger}}, \bibinfo {author} {\bibfnamefont {P.}~\bibnamefont
  {Bourgeois-Hope}}, \bibinfo {author} {\bibfnamefont {H.}~\bibnamefont
  {Taniguchi}}, \bibinfo {author} {\bibfnamefont {S.}~\bibnamefont {Ren\'e~de
  Cotret}}, \bibinfo {author} {\bibfnamefont {G.}~\bibnamefont
  {Grissonnanche}}, \bibinfo {author} {\bibfnamefont {M.~S.}\ \bibnamefont
  {Anwar}}, \bibinfo {author} {\bibfnamefont {Y.}~\bibnamefont {Maeno}},
  \bibinfo {author} {\bibfnamefont {N.}~\bibnamefont {Doiron-Leyraud}}, \ and\
  \bibinfo {author} {\bibfnamefont {L.}~\bibnamefont {Taillefer}},\ }\bibfield
  {title} {\enquote {\bibinfo {title} {Vertical line nodes in the
  superconducting gap structure of \ce{Sr2RuO4}},}\ }\href {\doibase
  10.1103/PhysRevX.7.011032} {\bibfield  {journal} {\bibinfo  {journal} {Phys.
  Rev. X}\ }\textbf {\bibinfo {volume} {7}},\ \bibinfo {pages} {011032}
  (\bibinfo {year} {2017})}\BibitemShut {NoStop}%
\bibitem [{\citenamefont {Kittaka}\ \emph {et~al.}(2018)\citenamefont
  {Kittaka}, \citenamefont {Nakamura}, \citenamefont {Sakakibara},
  \citenamefont {Kikugawa}, \citenamefont {Terashima}, \citenamefont {Uji},
  \citenamefont {Sokolov}, \citenamefont {Mackenzie}, \citenamefont {Irie},
  \citenamefont {Tsutsumi}, \citenamefont {Suzuki},\ and\ \citenamefont
  {Machida}}]{Kittaka:2018}%
  \BibitemOpen
  \bibfield  {author} {\bibinfo {author} {\bibfnamefont {S.}~\bibnamefont
  {Kittaka}}, \bibinfo {author} {\bibfnamefont {S.}~\bibnamefont {Nakamura}},
  \bibinfo {author} {\bibfnamefont {T.}~\bibnamefont {Sakakibara}}, \bibinfo
  {author} {\bibfnamefont {N.}~\bibnamefont {Kikugawa}}, \bibinfo {author}
  {\bibfnamefont {T.}~\bibnamefont {Terashima}}, \bibinfo {author}
  {\bibfnamefont {S.}~\bibnamefont {Uji}}, \bibinfo {author} {\bibfnamefont
  {D.~A.}\ \bibnamefont {Sokolov}}, \bibinfo {author} {\bibfnamefont {A.~P.}\
  \bibnamefont {Mackenzie}}, \bibinfo {author} {\bibfnamefont {K.}~\bibnamefont
  {Irie}}, \bibinfo {author} {\bibfnamefont {Y.}~\bibnamefont {Tsutsumi}},
  \bibinfo {author} {\bibfnamefont {K.}~\bibnamefont {Suzuki}}, \ and\ \bibinfo
  {author} {\bibfnamefont {K.}~\bibnamefont {Machida}},\ }\bibfield  {title}
  {\enquote {\bibinfo {title} {Searching for gap zeros in \ce{Sr2RuO4} via
  field-angle-dependent specific-heat measurement},}\ }\href {\doibase
  10.7566/JPSJ.87.093703} {\bibfield  {journal} {\bibinfo  {journal} {J.\
  Phys.\ Soc.\ Jpn.}\ }\textbf {\bibinfo {volume} {87}},\ \bibinfo {pages}
  {093703} (\bibinfo {year} {2018})}\BibitemShut {NoStop}%
\bibitem [{\citenamefont {Yonezawa}\ \emph {et~al.}(2013)\citenamefont
  {Yonezawa}, \citenamefont {Kajikawa},\ and\ \citenamefont
  {Maeno}}]{Yonezawa:2013}%
  \BibitemOpen
  \bibfield  {author} {\bibinfo {author} {\bibfnamefont {S.}~\bibnamefont
  {Yonezawa}}, \bibinfo {author} {\bibfnamefont {T.}~\bibnamefont {Kajikawa}},
  \ and\ \bibinfo {author} {\bibfnamefont {Y.}~\bibnamefont {Maeno}},\
  }\bibfield  {title} {\enquote {\bibinfo {title} {First-order superconducting
  transition of \ce{Sr2RuO4}},}\ }\href {\doibase
  10.1103/PhysRevLett.110.077003} {\bibfield  {journal} {\bibinfo  {journal}
  {Phys.\ Rev.\ Lett.}\ }\textbf {\bibinfo {volume} {110}},\ \bibinfo {pages}
  {077003} (\bibinfo {year} {2013})}\BibitemShut {NoStop}%
\bibitem [{\citenamefont {Kittaka}\ \emph {et~al.}(2014)\citenamefont
  {Kittaka}, \citenamefont {Kasahara}, \citenamefont {Sakakibara},
  \citenamefont {Shibata}, \citenamefont {Yonezawa}, \citenamefont {Maeno},
  \citenamefont {Tenya},\ and\ \citenamefont {Machida}}]{Kittaka:2014}%
  \BibitemOpen
  \bibfield  {author} {\bibinfo {author} {\bibfnamefont {S.}~\bibnamefont
  {Kittaka}}, \bibinfo {author} {\bibfnamefont {A.}~\bibnamefont {Kasahara}},
  \bibinfo {author} {\bibfnamefont {T.}~\bibnamefont {Sakakibara}}, \bibinfo
  {author} {\bibfnamefont {D.}~\bibnamefont {Shibata}}, \bibinfo {author}
  {\bibfnamefont {S.}~\bibnamefont {Yonezawa}}, \bibinfo {author}
  {\bibfnamefont {Y.}~\bibnamefont {Maeno}}, \bibinfo {author} {\bibfnamefont
  {K.}~\bibnamefont {Tenya}}, \ and\ \bibinfo {author} {\bibfnamefont
  {K.}~\bibnamefont {Machida}},\ }\bibfield  {title} {\enquote {\bibinfo
  {title} {Sharp magnetization jump at the first-order superconducting
  transition in \ce{Sr2RuO4}},}\ }\href {\doibase 10.1103/PhysRevB.90.220502}
  {\bibfield  {journal} {\bibinfo  {journal} {Phys.\ Rev.\ B}\ }\textbf
  {\bibinfo {volume} {90}},\ \bibinfo {pages} {220502} (\bibinfo {year}
  {2014})}\BibitemShut {NoStop}%
\bibitem [{\citenamefont {Kikugawa}\ \emph {et~al.}(2016)\citenamefont
  {Kikugawa}, \citenamefont {Terashima}, \citenamefont {Uji}, \citenamefont
  {Sugii}, \citenamefont {Maeno}, \citenamefont {Graf}, \citenamefont
  {Baumbach},\ and\ \citenamefont {Brooks}}]{Kikugawa:2016}%
  \BibitemOpen
  \bibfield  {author} {\bibinfo {author} {\bibfnamefont {N.}~\bibnamefont
  {Kikugawa}}, \bibinfo {author} {\bibfnamefont {T.}~\bibnamefont {Terashima}},
  \bibinfo {author} {\bibfnamefont {S.}~\bibnamefont {Uji}}, \bibinfo {author}
  {\bibfnamefont {K.}~\bibnamefont {Sugii}}, \bibinfo {author} {\bibfnamefont
  {Y.}~\bibnamefont {Maeno}}, \bibinfo {author} {\bibfnamefont
  {D.}~\bibnamefont {Graf}}, \bibinfo {author} {\bibfnamefont {R.}~\bibnamefont
  {Baumbach}}, \ and\ \bibinfo {author} {\bibfnamefont {J.}~\bibnamefont
  {Brooks}},\ }\bibfield  {title} {\enquote {\bibinfo {title} {Superconducting
  subphase in the layered perovskite ruthenate \ce{Sr2RuO4} in a parallel
  magnetic field},}\ }\href {\doibase 10.1103/PhysRevB.93.184513} {\bibfield
  {journal} {\bibinfo  {journal} {Phys.\ Rev.\ B}\ }\textbf {\bibinfo {volume}
  {93}},\ \bibinfo {pages} {184513} (\bibinfo {year} {2016})}\BibitemShut
  {NoStop}%
\bibitem [{\citenamefont {Ishida}\ \emph {et~al.}(1998)\citenamefont {Ishida},
  \citenamefont {Mukuda}, \citenamefont {Kitaoka}, \citenamefont {Asayama},
  \citenamefont {Mao}, \citenamefont {Mori},\ and\ \citenamefont
  {Maeno}}]{Ishida:1998}%
  \BibitemOpen
  \bibfield  {author} {\bibinfo {author} {\bibfnamefont {K.}~\bibnamefont
  {Ishida}}, \bibinfo {author} {\bibfnamefont {H.}~\bibnamefont {Mukuda}},
  \bibinfo {author} {\bibfnamefont {Y.}~\bibnamefont {Kitaoka}}, \bibinfo
  {author} {\bibfnamefont {K.}~\bibnamefont {Asayama}}, \bibinfo {author}
  {\bibfnamefont {Z.~Q.}\ \bibnamefont {Mao}}, \bibinfo {author} {\bibfnamefont
  {Y.}~\bibnamefont {Mori}}, \ and\ \bibinfo {author} {\bibfnamefont
  {Y.}~\bibnamefont {Maeno}},\ }\bibfield  {title} {\enquote {\bibinfo {title}
  {Spin-triplet superconductivity in \ce{Sr2RuO4} identified by
  $^{17}\mathrm{O}$ knight shift},}\ }\href {\doibase 10.1038/25315} {\bibfield
   {journal} {\bibinfo  {journal} {Nature}\ }\textbf {\bibinfo {volume}
  {396}},\ \bibinfo {pages} {658--660} (\bibinfo {year} {1998})}\BibitemShut
  {NoStop}%
\bibitem [{\citenamefont {Ishida}\ \emph {et~al.}(2000)\citenamefont {Ishida},
  \citenamefont {Mukuda}, \citenamefont {Kitaoka}, \citenamefont {Mao},
  \citenamefont {Mori},\ and\ \citenamefont {Maeno}}]{Ishida:2000}%
  \BibitemOpen
  \bibfield  {author} {\bibinfo {author} {\bibfnamefont {K.}~\bibnamefont
  {Ishida}}, \bibinfo {author} {\bibfnamefont {H.}~\bibnamefont {Mukuda}},
  \bibinfo {author} {\bibfnamefont {Y.}~\bibnamefont {Kitaoka}}, \bibinfo
  {author} {\bibfnamefont {Z.~Q.}\ \bibnamefont {Mao}}, \bibinfo {author}
  {\bibfnamefont {Y.}~\bibnamefont {Mori}}, \ and\ \bibinfo {author}
  {\bibfnamefont {Y.}~\bibnamefont {Maeno}},\ }\bibfield  {title} {\enquote
  {\bibinfo {title} {Anisotropic superconducting gap in the spin-triplet
  superconductor \ce{Sr2RuO4}: {E}vidence from a {Ru-NQR} study},}\ }\href
  {\doibase 10.1103/PhysRevLett.84.5387} {\bibfield  {journal} {\bibinfo
  {journal} {Phys.\ Rev.\ Lett.}\ }\textbf {\bibinfo {volume} {84}},\ \bibinfo
  {pages} {5387--5390} (\bibinfo {year} {2000})}\BibitemShut {NoStop}%
\bibitem [{\citenamefont {Murakawa}\ \emph {et~al.}(2004)\citenamefont
  {Murakawa}, \citenamefont {Ishida}, \citenamefont {Kitagawa}, \citenamefont
  {Mao},\ and\ \citenamefont {Maeno}}]{Murakawa:2004}%
  \BibitemOpen
  \bibfield  {author} {\bibinfo {author} {\bibfnamefont {H.}~\bibnamefont
  {Murakawa}}, \bibinfo {author} {\bibfnamefont {K.}~\bibnamefont {Ishida}},
  \bibinfo {author} {\bibfnamefont {K.}~\bibnamefont {Kitagawa}}, \bibinfo
  {author} {\bibfnamefont {Z.~Q.}\ \bibnamefont {Mao}}, \ and\ \bibinfo
  {author} {\bibfnamefont {Y.}~\bibnamefont {Maeno}},\ }\bibfield  {title}
  {\enquote {\bibinfo {title} {Measurement of the
  $^{101}\mathrm{R}\mathrm{u}$-knight shift of superconducting \ce{Sr2RuO4} in
  a parallel magnetic field},}\ }\href {\doibase 10.1103/PhysRevLett.93.167004}
  {\bibfield  {journal} {\bibinfo  {journal} {Phys.\ Rev.\ Lett.}\ }\textbf
  {\bibinfo {volume} {93}},\ \bibinfo {pages} {167004} (\bibinfo {year}
  {2004})}\BibitemShut {NoStop}%
\bibitem [{\citenamefont {Duffy}\ \emph {et~al.}(2000)\citenamefont {Duffy},
  \citenamefont {Hayden}, \citenamefont {Maeno}, \citenamefont {Mao},
  \citenamefont {Kulda},\ and\ \citenamefont {McIntyre}}]{Duffy:2000}%
  \BibitemOpen
  \bibfield  {author} {\bibinfo {author} {\bibfnamefont {J.~A.}\ \bibnamefont
  {Duffy}}, \bibinfo {author} {\bibfnamefont {S.~M.}\ \bibnamefont {Hayden}},
  \bibinfo {author} {\bibfnamefont {Y.}~\bibnamefont {Maeno}}, \bibinfo
  {author} {\bibfnamefont {Z.}~\bibnamefont {Mao}}, \bibinfo {author}
  {\bibfnamefont {J.}~\bibnamefont {Kulda}}, \ and\ \bibinfo {author}
  {\bibfnamefont {G.~J.}\ \bibnamefont {McIntyre}},\ }\bibfield  {title}
  {\enquote {\bibinfo {title} {Polarized-neutron scattering study of the
  cooper-pair moment in \ce{Sr2RuO4}},}\ }\href {\doibase
  10.1103/PhysRevLett.85.5412} {\bibfield  {journal} {\bibinfo  {journal}
  {Phys.\ Rev.\ Lett.}\ }\textbf {\bibinfo {volume} {85}},\ \bibinfo {pages}
  {5412--5415} (\bibinfo {year} {2000})}\BibitemShut {NoStop}%
\bibitem [{\citenamefont {Ishida}\ \emph {et~al.}(2015)\citenamefont {Ishida},
  \citenamefont {Manago}, \citenamefont {Yamanaka}, \citenamefont {Fukazawa},
  \citenamefont {Mao}, \citenamefont {Maeno},\ and\ \citenamefont
  {Miyake}}]{Ishida:2015}%
  \BibitemOpen
  \bibfield  {author} {\bibinfo {author} {\bibfnamefont {K.}~\bibnamefont
  {Ishida}}, \bibinfo {author} {\bibfnamefont {M.}~\bibnamefont {Manago}},
  \bibinfo {author} {\bibfnamefont {T.}~\bibnamefont {Yamanaka}}, \bibinfo
  {author} {\bibfnamefont {H.}~\bibnamefont {Fukazawa}}, \bibinfo {author}
  {\bibfnamefont {Z.~Q.}\ \bibnamefont {Mao}}, \bibinfo {author} {\bibfnamefont
  {Y.}~\bibnamefont {Maeno}}, \ and\ \bibinfo {author} {\bibfnamefont
  {K.}~\bibnamefont {Miyake}},\ }\bibfield  {title} {\enquote {\bibinfo {title}
  {Spin polarization enhanced by spin-triplet pairing in \ce{Sr2RuO4} probed by
  {NMR}},}\ }\href {\doibase 10.1103/PhysRevB.92.100502} {\bibfield  {journal}
  {\bibinfo  {journal} {Phys.\ Rev.\ B}\ }\textbf {\bibinfo {volume} {92}},\
  \bibinfo {pages} {100502} (\bibinfo {year} {2015})}\BibitemShut {NoStop}%
\bibitem [{\citenamefont {Hicks}\ \emph {et~al.}(2014)\citenamefont {Hicks},
  \citenamefont {Brodsky}, \citenamefont {Yelland}, \citenamefont {Gibbs},
  \citenamefont {Bruin}, \citenamefont {Barber}, \citenamefont {Edkins},
  \citenamefont {Nishimura}, \citenamefont {Yonezawa}, \citenamefont {Maeno},\
  and\ \citenamefont {Mackenzie}}]{Hicks:2014}%
  \BibitemOpen
  \bibfield  {author} {\bibinfo {author} {\bibfnamefont {C.~W.}\ \bibnamefont
  {Hicks}}, \bibinfo {author} {\bibfnamefont {D.~O.}\ \bibnamefont {Brodsky}},
  \bibinfo {author} {\bibfnamefont {E.~A.}\ \bibnamefont {Yelland}}, \bibinfo
  {author} {\bibfnamefont {A.~S.}\ \bibnamefont {Gibbs}}, \bibinfo {author}
  {\bibfnamefont {J.~A.~N.}\ \bibnamefont {Bruin}}, \bibinfo {author}
  {\bibfnamefont {M.~E.}\ \bibnamefont {Barber}}, \bibinfo {author}
  {\bibfnamefont {S.~D.}\ \bibnamefont {Edkins}}, \bibinfo {author}
  {\bibfnamefont {K.}~\bibnamefont {Nishimura}}, \bibinfo {author}
  {\bibfnamefont {S.}~\bibnamefont {Yonezawa}}, \bibinfo {author}
  {\bibfnamefont {Y.}~\bibnamefont {Maeno}}, \ and\ \bibinfo {author}
  {\bibfnamefont {A.~P.}\ \bibnamefont {Mackenzie}},\ }\bibfield  {title}
  {\enquote {\bibinfo {title} {Strong increase of ${T}_\mathrm{c}$ of
  \ce{Sr2RuO4} under both tensile and compressive strain},}\ }\href {\doibase
  10.1126/science.1248292} {\bibfield  {journal} {\bibinfo  {journal}
  {Science}\ }\textbf {\bibinfo {volume} {344}},\ \bibinfo {pages} {283--285}
  (\bibinfo {year} {2014})}\BibitemShut {NoStop}%
\bibitem [{\citenamefont {Steppke}\ \emph {et~al.}(2017)\citenamefont
  {Steppke}, \citenamefont {Zhao}, \citenamefont {Barber}, \citenamefont
  {Scaffidi}, \citenamefont {Jerzembeck}, \citenamefont {Rosner}, \citenamefont
  {Gibbs}, \citenamefont {Maeno}, \citenamefont {Simon}, \citenamefont
  {Mackenzie},\ and\ \citenamefont {Hicks}}]{Steppke:2017}%
  \BibitemOpen
  \bibfield  {author} {\bibinfo {author} {\bibfnamefont {A.}~\bibnamefont
  {Steppke}}, \bibinfo {author} {\bibfnamefont {L.}~\bibnamefont {Zhao}},
  \bibinfo {author} {\bibfnamefont {M.~E.}\ \bibnamefont {Barber}}, \bibinfo
  {author} {\bibfnamefont {T.}~\bibnamefont {Scaffidi}}, \bibinfo {author}
  {\bibfnamefont {F.}~\bibnamefont {Jerzembeck}}, \bibinfo {author}
  {\bibfnamefont {H.}~\bibnamefont {Rosner}}, \bibinfo {author} {\bibfnamefont
  {A.~S.}\ \bibnamefont {Gibbs}}, \bibinfo {author} {\bibfnamefont
  {Y.}~\bibnamefont {Maeno}}, \bibinfo {author} {\bibfnamefont {S.~H.}\
  \bibnamefont {Simon}}, \bibinfo {author} {\bibfnamefont {A.~P.}\ \bibnamefont
  {Mackenzie}}, \ and\ \bibinfo {author} {\bibfnamefont {C.~W.}\ \bibnamefont
  {Hicks}},\ }\bibfield  {title} {\enquote {\bibinfo {title} {Strong peak in
  ${T}_\mathrm{c}$ of \ce{Sr2RuO4} under uniaxial pressure},}\ }\href {\doibase
  10.1126/science.aaf9398} {\bibfield  {journal} {\bibinfo  {journal}
  {Science}\ }\textbf {\bibinfo {volume} {355}} (\bibinfo {year} {2017}),\
  10.1126/science.aaf9398}\BibitemShut {NoStop}%
\bibitem [{\citenamefont {Barber}\ \emph {et~al.}(2018)\citenamefont {Barber},
  \citenamefont {Gibbs}, \citenamefont {Maeno}, \citenamefont {Mackenzie},\
  and\ \citenamefont {Hicks}}]{Barber:2018}%
  \BibitemOpen
  \bibfield  {author} {\bibinfo {author} {\bibfnamefont {M.~E.}\ \bibnamefont
  {Barber}}, \bibinfo {author} {\bibfnamefont {A.~S.}\ \bibnamefont {Gibbs}},
  \bibinfo {author} {\bibfnamefont {Y.}~\bibnamefont {Maeno}}, \bibinfo
  {author} {\bibfnamefont {A.~P.}\ \bibnamefont {Mackenzie}}, \ and\ \bibinfo
  {author} {\bibfnamefont {C.~W.}\ \bibnamefont {Hicks}},\ }\bibfield  {title}
  {\enquote {\bibinfo {title} {Resistivity in the vicinity of a van {H}ove
  singularity: \ce{Sr2RuO4} under uniaxial pressure},}\ }\href {\doibase
  10.1103/PhysRevLett.120.076602} {\bibfield  {journal} {\bibinfo  {journal}
  {Phys.\ Rev.\ Lett.}\ }\textbf {\bibinfo {volume} {120}},\ \bibinfo {pages}
  {076602} (\bibinfo {year} {2018})}\BibitemShut {NoStop}%
\bibitem [{\citenamefont {Luo}\ \emph {et~al.}()\citenamefont {Luo},
  \citenamefont {Guzman}, \citenamefont {Dioguardi}, \citenamefont {Pustogow},
  \citenamefont {Thomas}, \citenamefont {Ronning}, \citenamefont {Kikugawa},
  \citenamefont {Sokolov}, \citenamefont {Jerzembeck}, \citenamefont
  {Mackenzie}, \citenamefont {Hicks}, \citenamefont {Bauer}, \citenamefont
  {Mazin},\ and\ \citenamefont {Brown}}]{Luo:arXiv}%
  \BibitemOpen
  \bibfield  {author} {\bibinfo {author} {\bibfnamefont {Y.}~\bibnamefont
  {Luo}}, \bibinfo {author} {\bibfnamefont {P.}~\bibnamefont {Guzman}},
  \bibinfo {author} {\bibfnamefont {A.~P.}\ \bibnamefont {Dioguardi}}, \bibinfo
  {author} {\bibfnamefont {A.}~\bibnamefont {Pustogow}}, \bibinfo {author}
  {\bibfnamefont {S.~M.}\ \bibnamefont {Thomas}}, \bibinfo {author}
  {\bibfnamefont {F.}~\bibnamefont {Ronning}}, \bibinfo {author} {\bibfnamefont
  {N.}~\bibnamefont {Kikugawa}}, \bibinfo {author} {\bibfnamefont
  {D.}~\bibnamefont {Sokolov}}, \bibinfo {author} {\bibfnamefont
  {F.}~\bibnamefont {Jerzembeck}}, \bibinfo {author} {\bibfnamefont {A.~P.}\
  \bibnamefont {Mackenzie}}, \bibinfo {author} {\bibfnamefont {C.~W.}\
  \bibnamefont {Hicks}}, \bibinfo {author} {\bibfnamefont {E.~D.}\ \bibnamefont
  {Bauer}}, \bibinfo {author} {\bibfnamefont {I.~I.}\ \bibnamefont {Mazin}}, \
  and\ \bibinfo {author} {\bibfnamefont {S.~E.}\ \bibnamefont {Brown}},\
  }\bibfield  {title} {\enquote {\bibinfo {title} {Normal state
  $^{17}\mathrm{O}$ {NMR} studies of \ce{Sr2RuO4} under uniaxial stress},}\
  }\href {https://arxiv.org/abs/1810.01209} {\bibfield  {journal} {\bibinfo
  {journal} {{\it arXiv:1810.01209v1}}\ }}\bibinfo {note}
  {{h}ttps://arxiv.org/abs/1810.01209}\BibitemShut {NoStop}%
\bibitem [{\citenamefont {Maeno}\ \emph {et~al.}(1998)\citenamefont {Maeno},
  \citenamefont {Ando}, \citenamefont {Mori}, \citenamefont {Ohmichi},
  \citenamefont {Ikeda}, \citenamefont {NishiZaki},\ and\ \citenamefont
  {Nakatsuji}}]{Maeno:1998}%
  \BibitemOpen
  \bibfield  {author} {\bibinfo {author} {\bibfnamefont {Y.}~\bibnamefont
  {Maeno}}, \bibinfo {author} {\bibfnamefont {T.}~\bibnamefont {Ando}},
  \bibinfo {author} {\bibfnamefont {Y.}~\bibnamefont {Mori}}, \bibinfo {author}
  {\bibfnamefont {E.}~\bibnamefont {Ohmichi}}, \bibinfo {author} {\bibfnamefont
  {S.}~\bibnamefont {Ikeda}}, \bibinfo {author} {\bibfnamefont
  {S.}~\bibnamefont {NishiZaki}}, \ and\ \bibinfo {author} {\bibfnamefont
  {S.}~\bibnamefont {Nakatsuji}},\ }\bibfield  {title} {\enquote {\bibinfo
  {title} {Enhancement of superconductivity of \ce{Sr2RuO4} to 3~{K} by
  embedded metallic microdomains},}\ }\href {\doibase
  10.1103/PhysRevLett.81.3765} {\bibfield  {journal} {\bibinfo  {journal}
  {Phys.\ Rev.\ Lett.}\ }\textbf {\bibinfo {volume} {81}},\ \bibinfo {pages}
  {3765--3768} (\bibinfo {year} {1998})}\BibitemShut {NoStop}%
\bibitem [{\citenamefont {Ando}\ \emph {et~al.}(1999)\citenamefont {Ando},
  \citenamefont {Akima}, \citenamefont {Mori},\ and\ \citenamefont
  {Maeno}}]{Ando:1999}%
  \BibitemOpen
  \bibfield  {author} {\bibinfo {author} {\bibfnamefont {T.}~\bibnamefont
  {Ando}}, \bibinfo {author} {\bibfnamefont {T.}~\bibnamefont {Akima}},
  \bibinfo {author} {\bibfnamefont {Y.}~\bibnamefont {Mori}}, \ and\ \bibinfo
  {author} {\bibfnamefont {Y.}~\bibnamefont {Maeno}},\ }\bibfield  {title}
  {\enquote {\bibinfo {title} {Upper critical fields of the 3-{K}
  superconducting phase of \ce{Sr2RuO4}},}\ }\href {\doibase
  10.1143/JPSJ.68.1651} {\bibfield  {journal} {\bibinfo  {journal} {J.\ Phys.\
  Soc.\ Jpn.}\ }\textbf {\bibinfo {volume} {68}},\ \bibinfo {pages}
  {1651--1656} (\bibinfo {year} {1999})}\BibitemShut {NoStop}%
\bibitem [{\citenamefont {Yaguchi}\ \emph {et~al.}(2003)\citenamefont
  {Yaguchi}, \citenamefont {Wada}, \citenamefont {Akima}, \citenamefont
  {Maeno},\ and\ \citenamefont {Ishiguro}}]{Yaguchi:2003}%
  \BibitemOpen
  \bibfield  {author} {\bibinfo {author} {\bibfnamefont {H.}~\bibnamefont
  {Yaguchi}}, \bibinfo {author} {\bibfnamefont {M.}~\bibnamefont {Wada}},
  \bibinfo {author} {\bibfnamefont {T.}~\bibnamefont {Akima}}, \bibinfo
  {author} {\bibfnamefont {Y.}~\bibnamefont {Maeno}}, \ and\ \bibinfo {author}
  {\bibfnamefont {T.}~\bibnamefont {Ishiguro}},\ }\bibfield  {title} {\enquote
  {\bibinfo {title} {Interface superconductivity in the eutectic
  \ce{Sr2RuO4}\ensuremath{-}\ce{Ru}: 3-{K} phase of \ce{Sr2RuO4}},}\ }\href
  {\doibase 10.1103/PhysRevB.67.214519} {\bibfield  {journal} {\bibinfo
  {journal} {Phys.\ Rev.\ B}\ }\textbf {\bibinfo {volume} {67}},\ \bibinfo
  {pages} {214519} (\bibinfo {year} {2003})}\BibitemShut {NoStop}%
\bibitem [{\citenamefont {Mao}\ \emph {et~al.}(2000)\citenamefont {Mao},
  \citenamefont {Fukazawa},\ and\ \citenamefont {Maeno}}]{Mao:2000}%
  \BibitemOpen
  \bibfield  {author} {\bibinfo {author} {\bibfnamefont {Z.~Q.}\ \bibnamefont
  {Mao}}, \bibinfo {author} {\bibfnamefont {H.}~\bibnamefont {Fukazawa}}, \
  and\ \bibinfo {author} {\bibfnamefont {Y.}~\bibnamefont {Maeno}},\ }\bibfield
   {title} {\enquote {\bibinfo {title} {Crystal growth of \ce{Sr2RuO4}},}\
  }\href {\doibase 10.1016/S0025-5408(00)00378-0} {\bibfield  {journal}
  {\bibinfo  {journal} {Mater.\ Res.\ Bull.}\ }\textbf {\bibinfo {volume}
  {35}},\ \bibinfo {pages} {1813--1824} (\bibinfo {year} {2000})}\BibitemShut
  {NoStop}%
\bibitem [{\citenamefont {Perry}\ and\ \citenamefont
  {Maeno}(2004)}]{Perry:2004}%
  \BibitemOpen
  \bibfield  {author} {\bibinfo {author} {\bibfnamefont {R.~S.}\ \bibnamefont
  {Perry}}\ and\ \bibinfo {author} {\bibfnamefont {Y.}~\bibnamefont {Maeno}},\
  }\bibfield  {title} {\enquote {\bibinfo {title} {Systematic approach to the
  growth of high-quality single crystals of \ce{Sr3Ru2O7}},}\ }\href {\doibase
  10.1016/j.jcrysgro.2004.07.082} {\bibfield  {journal} {\bibinfo  {journal}
  {J.\ Cryst.\ Growth}\ }\textbf {\bibinfo {volume} {271}},\ \bibinfo {pages}
  {942--950} (\bibinfo {year} {2004})}\BibitemShut {NoStop}%
\bibitem [{\citenamefont {Zhou}\ \emph {et~al.}(2005)\citenamefont {Zhou},
  \citenamefont {Hooper}, \citenamefont {Fobes}, \citenamefont {Mao},
  \citenamefont {Golub},\ and\ \citenamefont {O'Connor}}]{Zhou:2005}%
  \BibitemOpen
  \bibfield  {author} {\bibinfo {author} {\bibfnamefont {M.}~\bibnamefont
  {Zhou}}, \bibinfo {author} {\bibfnamefont {J.}~\bibnamefont {Hooper}},
  \bibinfo {author} {\bibfnamefont {D.}~\bibnamefont {Fobes}}, \bibinfo
  {author} {\bibfnamefont {Z.~Q.}\ \bibnamefont {Mao}}, \bibinfo {author}
  {\bibfnamefont {V.}~\bibnamefont {Golub}}, \ and\ \bibinfo {author}
  {\bibfnamefont {C.~J.}\ \bibnamefont {O'Connor}},\ }\bibfield  {title}
  {\enquote {\bibinfo {title} {Electronic and magnetic properties of
  triple-layered ruthenate \ce{Sr4Ru3O10} single crystals grown by a
  floating-zone method},}\ }\href {\doibase 10.1016/j.materresbull.2005.03.004}
  {\bibfield  {journal} {\bibinfo  {journal} {Mater.\ Res.\ Bull.}\ }\textbf
  {\bibinfo {volume} {40}},\ \bibinfo {pages} {942--950} (\bibinfo {year}
  {2005})}\BibitemShut {NoStop}%
\bibitem [{\citenamefont {Kikugawa}\ \emph {et~al.}(2009)\citenamefont
  {Kikugawa}, \citenamefont {Balicas},\ and\ \citenamefont
  {Mackenzie}}]{Kikugawa:2009}%
  \BibitemOpen
  \bibfield  {author} {\bibinfo {author} {\bibfnamefont {N.}~\bibnamefont
  {Kikugawa}}, \bibinfo {author} {\bibfnamefont {L.}~\bibnamefont {Balicas}}, \
  and\ \bibinfo {author} {\bibfnamefont {A.~P.}\ \bibnamefont {Mackenzie}},\
  }\bibfield  {title} {\enquote {\bibinfo {title} {Physical properties of
  single-crystalline \ce{CaRuO3} grown by a floating-zone method},}\ }\href
  {\doibase 10.1143/JPSJ.78.014701} {\bibfield  {journal} {\bibinfo  {journal}
  {J.\ Phys.\ Soc.\ Jpn.}\ }\textbf {\bibinfo {volume} {78}},\ \bibinfo {pages}
  {014701} (\bibinfo {year} {2009})}\BibitemShut {NoStop}%
\bibitem [{\citenamefont {Kikugawa}\ \emph {et~al.}(2015)\citenamefont
  {Kikugawa}, \citenamefont {Baumbach}, \citenamefont {Brooks}, \citenamefont
  {Terashima}, \citenamefont {Uji},\ and\ \citenamefont
  {Maeno}}]{Kikugawa:2015}%
  \BibitemOpen
  \bibfield  {author} {\bibinfo {author} {\bibfnamefont {N.}~\bibnamefont
  {Kikugawa}}, \bibinfo {author} {\bibfnamefont {R.}~\bibnamefont {Baumbach}},
  \bibinfo {author} {\bibfnamefont {J.~S.}\ \bibnamefont {Brooks}}, \bibinfo
  {author} {\bibfnamefont {T.}~\bibnamefont {Terashima}}, \bibinfo {author}
  {\bibfnamefont {S.}~\bibnamefont {Uji}}, \ and\ \bibinfo {author}
  {\bibfnamefont {Y.}~\bibnamefont {Maeno}},\ }\bibfield  {title} {\enquote
  {\bibinfo {title} {Single-crystal growth of a perovskite ruthenate
  \ce{SrRuO3} by the floating-zone method},}\ }\href {\doibase
  10.1021/acs.cgd.5b01248} {\bibfield  {journal} {\bibinfo  {journal} {Cryst.\
  Growth Des.}\ }\textbf {\bibinfo {volume} {15}},\ \bibinfo {pages}
  {5573--5577} (\bibinfo {year} {2015})}\BibitemShut {NoStop}%
\bibitem [{\citenamefont {Yoshida}\ \emph {et~al.}(2004)\citenamefont
  {Yoshida}, \citenamefont {Nagai}, \citenamefont {Ikeda}, \citenamefont
  {Shirakawa}, \citenamefont {Kosaka},\ and\ \citenamefont
  {M\^ori}}]{Yoshida:2004}%
  \BibitemOpen
  \bibfield  {author} {\bibinfo {author} {\bibfnamefont {Y.}~\bibnamefont
  {Yoshida}}, \bibinfo {author} {\bibfnamefont {I.}~\bibnamefont {Nagai}},
  \bibinfo {author} {\bibfnamefont {S.-I.}\ \bibnamefont {Ikeda}}, \bibinfo
  {author} {\bibfnamefont {N.}~\bibnamefont {Shirakawa}}, \bibinfo {author}
  {\bibfnamefont {M.}~\bibnamefont {Kosaka}}, \ and\ \bibinfo {author}
  {\bibfnamefont {N.}~\bibnamefont {M\^ori}},\ }\bibfield  {title} {\enquote
  {\bibinfo {title} {Quasi-two-dimensional metallic ground state of
  \ce{Ca3Ru2O7}},}\ }\href {\doibase 10.1103/PhysRevB.69.220411} {\bibfield
  {journal} {\bibinfo  {journal} {Phys.\ Rev.\ B}\ }\textbf {\bibinfo {volume}
  {69}},\ \bibinfo {pages} {220411} (\bibinfo {year} {2004})}\BibitemShut
  {NoStop}%
\bibitem [{\citenamefont {Kikugawa}\ \emph {et~al.}(2010)\citenamefont
  {Kikugawa}, \citenamefont {Rost}, \citenamefont {Hicks}, \citenamefont
  {Schofield},\ and\ \citenamefont {Mackenzie}}]{Kikugawa:2010}%
  \BibitemOpen
  \bibfield  {author} {\bibinfo {author} {\bibfnamefont {N.}~\bibnamefont
  {Kikugawa}}, \bibinfo {author} {\bibfnamefont {A.~W.}\ \bibnamefont {Rost}},
  \bibinfo {author} {\bibfnamefont {C.~W.}\ \bibnamefont {Hicks}}, \bibinfo
  {author} {\bibfnamefont {A.~J.}\ \bibnamefont {Schofield}}, \ and\ \bibinfo
  {author} {\bibfnamefont {A.~P.}\ \bibnamefont {Mackenzie}},\ }\bibfield
  {title} {\enquote {\bibinfo {title} {\ce{Ca3Ru2O7}: {D}ensity wave formation
  and quantum oscillations in the hall resistivity},}\ }\href {\doibase
  10.1143/JPSJ.79.024704} {\bibfield  {journal} {\bibinfo  {journal} {J.\
  Phys.\ Soc.\ Jpn.}\ }\textbf {\bibinfo {volume} {79}},\ \bibinfo {pages}
  {024704} (\bibinfo {year} {2010})}\BibitemShut {NoStop}%
\bibitem [{\citenamefont {Yonezawa}\ \emph {et~al.}(2015)\citenamefont
  {Yonezawa}, \citenamefont {Higuchi}, \citenamefont {Sugimoto}, \citenamefont
  {Sow},\ and\ \citenamefont {Maeno}}]{Yonezawa:2015}%
  \BibitemOpen
  \bibfield  {author} {\bibinfo {author} {\bibfnamefont {S.}~\bibnamefont
  {Yonezawa}}, \bibinfo {author} {\bibfnamefont {T.}~\bibnamefont {Higuchi}},
  \bibinfo {author} {\bibfnamefont {Y.}~\bibnamefont {Sugimoto}}, \bibinfo
  {author} {\bibfnamefont {C.}~\bibnamefont {Sow}}, \ and\ \bibinfo {author}
  {\bibfnamefont {Y.}~\bibnamefont {Maeno}},\ }\bibfield  {title} {\enquote
  {\bibinfo {title} {Compact {AC} susceptometer for fast sample
  characterization down to 0.1~{K}},}\ }\href {\doibase 10.1063/1.4929871}
  {\bibfield  {journal} {\bibinfo  {journal} {Rev.\ Sci.\ Instrum.}\ }\textbf
  {\bibinfo {volume} {86}},\ \bibinfo {pages} {093903} (\bibinfo {year}
  {2015})}\BibitemShut {NoStop}%
\end{thebibliography}%

\end{document}